\documentclass[a4paper,fleqn]{cas-dc}
\usepackage{enumitem}
\usepackage[authoryear]{natbib}
\usepackage[figuresright]{rotating}
\usepackage{moreverb}
\usepackage{amsmath,amssymb,amsfonts}
\usepackage{algorithmic}
\usepackage{url}
\usepackage{graphicx}
\usepackage{textcomp}
\usepackage{xcolor}
\usepackage{tcolorbox}
\usepackage{ragged2e}
\usepackage{balance}
\usepackage{colortbl}
\usepackage{multirow}
\usepackage{color}

\usepackage{CJKutf8}
\usepackage{siunitx}
\usepackage{rotfloat}
\usepackage{float}
\usepackage{lscape}
\usepackage{pifont}
\usepackage{amssymb,mathrsfs,amsmath}
\usepackage{adjustbox}
\def\tsc#1{\csdef{#1}{\textsc{\lowercase{#1}}\xspace}}
\tsc{WGM}
\tsc{QE}
\tsc{EP}
\tsc{PMS}
\tsc{BEC}
\tsc{DE}

\newcommand{\cmark}{\ding{51}}
\newcommand{\xmark}{\ding{55}}

\begin{document}
\begin{sloppypar}

\let\WriteBookmarks\relax
\def\floatpagepagefraction{1}
\def\textpagefraction{.001}
\shorttitle{Issues, Causes and Solutions in Open-Source LLM-based Multi-Agent Systems}
\shortauthors{A U Rehman et al.}
\title [mode = title]{Understanding Issues, Causes and Solutions in Open-Source LLM-based Multi-Agent Systems}

\author[1]{Asad Ur Rehman}[orcid=0009-0005-8484-3287]
\ead{asadurrehman@whu.edu.cn}
\credit{Conceptualization, Investigation, Data curation, Formal analysis, Writing - Original draft preparation}

\author[1]{Syed Mohammad Kashif} [orcid=0009-0001-0243-0720]
\ead{smkashif145@gmail.com}
\credit{Investigation, Data curation, Writing - Original draft preparation}

\author[1]{Ruiyin Li} [orcid=0000-0001-8536-4935]
\cormark[1]
\ead{ryli_cs@whu.edu.cn}
\credit{Investigation, Data curation, Writing - Original draft preparation}

\author[1]{Peng Liang} [orcid=0000-0002-2056-5346]
\cormark[1]
\ead{liangp@whu.edu.cn}
\credit{Conceptualization, Methodology, Investigation, Data curation, Supervision, Writing - Original draft preparation}
\address[1]{School of Computer Science, Wuhan University, China}

\author[2]{Zengyang Li} [orcid=0000-0002-7258-993X]
\ead{zengyangli@ccnu.edu.cn}
\credit{Conceptualization, Methodology, Writing - review and editing}
\address[2]{School of Computer Science \& Hubei Provincial Key Laboratory of Artificial Intelligence and Smart Learning, Central China Normal University, China}

\author[3]{Arif Ali Khan} [orcid=0000-0002-8479-1481]
\ead{arif.khan@oulu.fi}
\credit{Conceptualization, Methodology, Writing - review and editing}
\address[3]{M3S Empirical Software Engineering Research Unit, University of Oulu, Finland}

\cortext[cor1]{Corresponding author.}

\begin{abstract}
With the advancement of LLM-based multi-agent systems (MAS), an increasing number of open-source projects are adopting multi-agent architectures as the foundation of their core functionality. Although research and practice on MAS have attracted considerable attention, limited studies have explored the challenges faced by practitioners of open-source LLM-based MAS, the causes of these challenges, and potential solutions. To address this gap, we conducted an empirical study to understand the issues that practitioners encounter when developing and using open-source LLM-based MAS, the possible causes of these issues, and potential solutions. We collected 22,848 closed issues from 21 open-source LLM-based MAS and applied a mixed automated and manual filtering approach to reduce the dataset to 944 issues related to LLM-based MAS. We then analyzed these issues to understand the frequent issues encountered by practitioners, their underlying causes, and potential solutions. Our study results show that (1) \emph{Orchestration \& Execution Issue} is the most common issue faced by practitioners, (2) \emph{Workflow Problem}, \emph{Tool Integration Problem}, and \emph{Memory Problem} are identified as the most frequent causes of the issues, and (3) \textit{Optimize Workflow} is the predominant solution to the issues. Based on the study results, we derive empirically grounded implications for practitioners and researchers aimed at improving orchestration, tool integration, and memory mechanisms in LLM-based MAS.
\end{abstract}

\begin{keywords}
LLM-based Multi-Agent System; Open Source Project; Issue; Cause; Solution
\end{keywords}

\maketitle

\section{Introduction}
\label{introduction}

Large Language Models (LLMs) are transforming the landscape of software development, moving us beyond simple coding copilots to a future where specialized LLM-powered agents actively collaborate like a digital team~\citep{hou2024large}. Since the release of ChatGPT in 2022 \citep{ChatGPT}, research on LLMs has grown rapidly~\citep{zhao2023survey}. The software engineering (SE) community quickly recognized their potential and began applying LLMs throughout the software development lifecycle~\citep{hou2024large,wang2024software}. Early studies examined single-agent scenarios, where an LLM assists developers with code generation, code review, software testing, and requirements documentation~\citep{fan2023large}. 

Recent advances in multi-agent frameworks have enabled the development of LLM-based multi-agent systems (MAS), in which multiple specialized LLM-powered agents communicate, coordinate, and collaborate to accomplish complex, sequential software engineering tasks that were previously too difficult for a single LLM agent~\citep{He2025MultiAgent, liu2026llmagentsSE}. This architectural approach is well-suited for complex end-to-end software processes because it addresses the inherent limitations of individual LLMs, including limited context windows, error propagation, and difficulties in managing complex software development workflows~\citep{He2025MultiAgent, wang2024software}.

Meanwhile, software practitioners are increasingly integrating LLM-based multi-agent systems (MAS) into their software projects. Frameworks such as LangChain and Semantic Kernel facilitate the development of these systems, supporting a wide range of applications from code generation to system design~\citep{Chase2024LangChain, Microsoft2024SemanticKernel}. Recent surveys have examined the use of LLMs in SE, analyzing model usage, dataset creation, and evaluation methods~\citep{hou2024large}. Despite the rapid adoption and significant potential of LLM-based MAS~\citep{Guo2024LLM}, there remains a lack of empirical studies regarding the challenges associated with their development and maintenance in real-world contexts. From an SE perspective, the complexity of agent coordination, planning, and interaction introduces novel challenges that are not yet fully understood.

This study addresses this gap by conducting a comprehensive empirical analysis of prominent open-source systems (OSS) that incorporate LLM-based MAS as core architectural components. We collected 22,848 closed issues from relevant GitHub repositories whose documentation explicitly describes them as MAS projects. GitHub is the leading platform for hosting OSS projects and is widely used in empirical SE research \citep{Alrashedy2024}. Our approach of analyzing closed issues from prominent OSS projects follows well-established practices in empirical SE \citep{waseem2026microservices, cai2025demystifying}. By analyzing issue tracking data, our study directly observes the practical issues and root causes that developers encounter, as well as the solutions they employ when building and maintaining LLM-based MAS.

\begin{figure}[htbp]
	\centering
	\includegraphics[width=\linewidth]{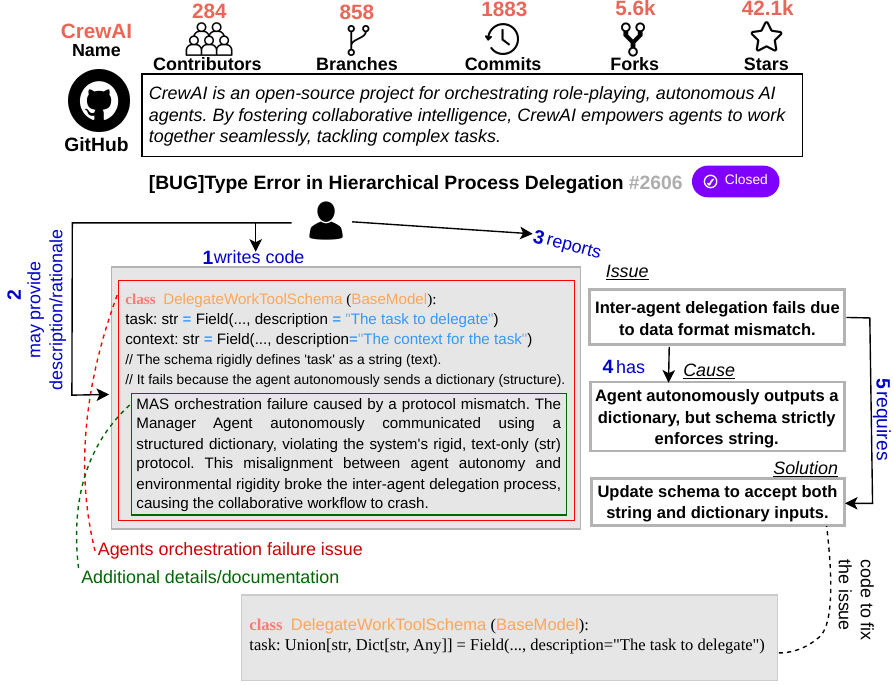}
	\caption{A GitHub issue demonstrating an \textit{Orchestration \& Execution Issue} in an LLM-based MAS.}
    \vspace{-0.3cm}
	\label{fig:mas-issue-example}
\end{figure}

\textbf{A motivating example:} Figure~\ref{fig:mas-issue-example} presents a sample GitHub issue (\href{https://github.com/crewAIInc/crewAI/issues/2606}{\texttt{CrewAI} issue \#2606}) from an LLM-based MAS project in which a developer reports a type validation issue where the manager agent passes structured dictionary inputs while the schema expects string values. To ensure consistency, the developer proposes enforcing a JSON message format, thereby highlighting an \textit{Orchestration \& Execution Issue}. Overall, the issue highlights a type mismatch in multi-agent coordination within LLM-based MAS.

Our study \textbf{aims} to identify the \textit{issues} in developing and using open-source LLM-based MAS and to determine the \textit{causes} of these issues, as well as possible \textit{solutions}. With this aim in mind, we formulate three Research Questions (RQs):

\begin{itemize}
    \item \textbf{RQ1}: What issues do practitioners face when working on LLM-based MAS?
    \item \textbf{RQ2}: What are the underlying causes of the issues in LLM-based MAS?
    \item \textbf{RQ3}: What are the potential solutions to address the issues in LLM-based MAS?
\end{itemize} 

To answer RQ1–RQ3, we analyzed the filtered set of 944 GitHub issues to identify the issues reported by practitioners, examine their underlying causes, and the solutions adopted in practice.

Our \textbf{findings} show that: (1) \textit{Orchestration \& Execution Issue} is the most common issue faced by practitioners, (2) \textit{Workflow Problem}, \textit{Tool Integration Problem}, and \textit{Memory Problem} are identified as the most frequent causes of the issues, and (3) \textit{Optimize Workflow} and \textit{Adjust Configuration} are the predominant solutions to the issues in LLM-based MAS.

The \textbf{main contributions} of this work are threefold: (1) We identify the key issues that occur during the development of LLM-based MAS projects using the dataset~\citep{replicationPackage} and organize them into a two-level taxonomy to provide a hierarchical understanding of their characteristics. (2) We identify the root causes of these issues and document the corresponding solutions proposed and adopted by practitioners, organizing both into taxonomies for causes and solutions. (3) We establish and analyze the mapping relationships between the identified issues and their underlying causes, as well as between the issues and their associated solutions, to better understand how issues are triggered and addressed in practice.

The rest of this paper is structured as follows: Section~\ref{sec:rqdataset} presents the Research Questions (RQs) and the research methodology employed in this study. Section~\ref{sec:ResultsandInterpretation} presents the study results. Section~\ref{sec:Implications} discusses the findings and implications based on the study results. Section~\ref{sec:ThreatstoValidity} discusses the threats to validity. Section~\ref{sec:relatedwork} reviews the related work, and Section~\ref{sec:Conclusions} concludes this work along with the directions for future research.

\section{Methodology}
\label{sec:rqdataset}
\begin{figure*}[htbp]
	\centering
	\includegraphics[width=1\textwidth]{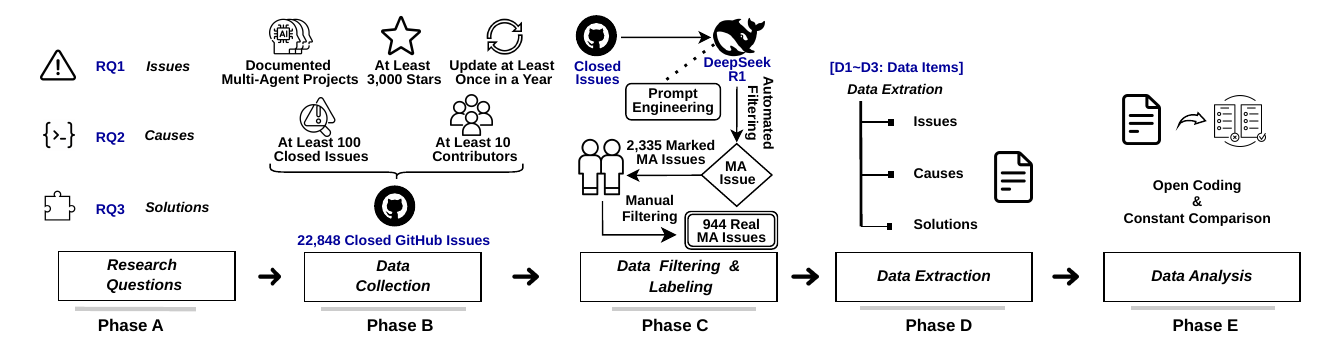}
	\caption{Overview of the research methodology}
	 \label{fig:methodology}
\end{figure*}

This study follows a systematic, multi-stage research methodology to investigate the practitioner-reported issues, underlying causes, and corresponding solutions in open-source LLM-based MAS. Based on this objective, we formulate three Research Questions (RQs) to guide the subsequent phases of the study and ensure a structured and reproducible analysis process, as shown in Figure~\ref{fig:methodology}. The research process comprises the formulation of RQs in Section~\ref{sec:goals}, data collection in Section~\ref{subsec:DataCollection}, data filtering and labeling in Section~\ref{subsec:DataFilteringandLabeling}, data extraction in Section~\ref{subsec:DataExtraction}, and data analysis in Section~\ref{subsec:DataAnalysis}.

\subsection{Research Questions}\label{sec:goals}

\textbf{\textit{RQ1:} What issues do practitioners face when working on MAS?}

\textit{Rationale}: As stated in Section~\ref{introduction}, recent advances in LLMs \citep{zhao2023survey} have led to a rapid increase in OSS that use LLMs as core components for building intelligent agents (i.e., LLM-based MAS). This RQ aims to provide a comprehensive understanding of the issues faced by OSS software engineers while developing MAS. This understanding provides an empirical foundation for identifying the critical challenges that require attention in the design and development of LLM-based MAS. Therefore, identifying and analyzing these practical difficulties is a fundamental first step in this study.

\textbf{\textit{RQ2:} What are the underlying causes of the issues in a MAS?}

\textit{Rationale}: Following the identification of issues currently faced by practitioners in \textit{RQ1}, it is necessary to determine their underlying causes. Identifying these root causes allows this study to provide more practical and actionable recommendations for developing effective, context-specific solutions. Additionally, these insights can inform the ongoing refinement and optimization of both LLM-based MAS and the underlying foundation models.

\textbf{\textit{RQ3:} What are the potential solutions to address the issues in MAS?}

\textit{Rationale}: This \textit{RQ} seeks to understand how OSS practitioners address the challenges encountered during the development and maintenance of LLM-based MAS. Answering this RQ will identify the most commonly adopted solutions within the MAS community and yield a diverse set of practical, practitioner-validated responses to the issues identified in \textit{RQ1}. By systematically exploring and analyzing these solutions, the study aims to provide actionable guidance for optimizing the design and implementation of LLM-based MAS.

\subsection{Data Collection}\label{subsec:DataCollection}
We collected data for our study from GitHub repositories that clearly mention MAS in their documentation. GitHub is currently the largest platform for hosting OSS projects. In addition, every project on GitHub has an “Issues” page, which provides a platform for users and developers of the project to report issues experienced during development and usage, and to discuss the identification, analysis, and resolution of these issues with other users and developers. In order to thoroughly explore the causes of issues and their potential solutions, we collected all ``closed'' issues from the repositories of the selected LLM-based MAS.

To ensure comprehensive data collection, we did not set a start date for the data source. The date we performed the data search was November 16, 2025, and we collected all the closed issues created before this date. The first author utilized a keyword-based search approach, using the search terms ``Multi-Agent System'', ``Multi-Agent LLM'', and ``AI Agents'' to search across GitHub. Any project that contained these search terms (e.g., in the project description or tags) was retrieved. After obtaining the search results, we found that although many projects include the terms ``Multi-Agent System'', ``Multi-Agent LLM'', or ``AI Agents'', they do not actually implement multi-agent architectures or utilize multiple LLM-based agents as core components. For example, some projects focus on teaching multi-agent concepts without building functional MAS. Therefore, we established the following criteria for selecting open-source LLM-Based MAS projects.

\begin{table}[htbp]
\caption{Inclusion and exclusion criteria for selecting LLM-based multi-agent open-source projects from GitHub repositories}
\setlength{\tabcolsep}{3pt}
\renewcommand{\arraystretch}{0.9}
\begin{tabular}{>{\raggedright\arraybackslash}p{0.05\linewidth} >{\raggedright\arraybackslash}p{0.85\linewidth}}
\toprule
\multicolumn{2}{l}{\textbf{Inclusion Criteria}} \\
\midrule
I1 & The project must have at least 3,000 stars.\\
I2 & The project must have at least 10 contributors.\\
I3 & The project must have at least 100 closed issues.\\
I4 & The project must have been updated at least once
within the past year.\\
I5 & The repository must be a real open-source software project that explicitly integrates multi-agent LLM architectures as core architectural components to execute its primary functionality. \\
\midrule
\multicolumn{2}{l}{\textbf{Exclusion Criteria}} \\
\midrule
E1 & We excluded projects categorized as course assignments, teaching guides, standalone LLM tutorials, books, or competition submissions to ensure the study focuses on practical software systems. \\
E2 & We also excluded repositories that merely mention multi-agent concepts in their documentation without implementing actual multi-agent functionality. \\
\bottomrule
\end{tabular}
\label{tab:inclusion-criteria}
\end{table}

Table~\ref{tab:inclusion-criteria} summarizes the inclusion and exclusion criteria used in this study. We defined an inclusion criterion (I5) to select projects that utilize multi-agent LLM architectures as core architectural components for their primary functionality. Following established empirical practices \citep{cai2025demystifying, lin2025empirical}, we further applied the inclusion criterion (I1)–(I4) as benchmarks for evaluating project quality, activity, and popularity. By enforcing thresholds on star counts, contributor participation, issue resolution history, and recent update activity, we excluded inactive or unpopular repositories. 

\begin{table}[htbp]
\scriptsize
\caption{Information of selected open-source LLM-based multi-agent projects}
\label{tbl:multi-agent-projects}
\begin{tabular}{>{\raggedright}m{2.4cm}m{0.71cm}<{\centering}m{1.32cm}<{\centering}m{0.71cm}<{\centering}m{1.5cm}<{\centering}}
\hline

\textbf{Project Name}       & \textbf{\#Stars} & \textbf{\#Issues}      & \textbf{\#Forks}    & \textbf{\#Contributors}    \\ \hline
Activepieces                & 19.1k            & 2216                   & 3.1k                & 336 \\ \hline
Adk-python                  & 14.2k            & 1354                   & 2.7k                & 180 \\ \hline
Agency-swarm                & 3.9k             & 141                    & 998                 & 21  \\ \hline
Agno                        & 34.8k            & 1475                   & 4.8k                & 343 \\ \hline
Ag2                         & 3.7k             & 646                    & 511                 & 158 \\ \hline
Agentscope                  & 13.5k            & 305                    & 1.2k                & 23  \\ \hline
Autogen                     & 55.6k            & 2496                   & 8.2k                & 547 \\ \hline
Camel                       & 14.7k            & 912                    & 1.7k                & 183 \\ \hline
ChatDev                     & 27.7k            & 255                    & 3.5k                & 56  \\ \hline
Continue                    & 29.9k            & 3384                   & 3.9k                & 421 \\ \hline
CrewAI                      & 39.9k            & 1548                   & 5.6k                & 278 \\ \hline
Claude-flow                 & 9.5k             & 311                    & 1.4k                & 11  \\ \hline
Flowise                     & 46.4k            & 1777                   & 23.4k               & 268 \\ \hline
Langroid                    & 3.7k             & 201                    & 348                 & 41  \\ \hline
MetaGPT                     & 59.2k            & 874                    & 7.6k                & 113 \\ \hline
Openai-agents-python        & 17k              & 811                    & 3k                  & 182 \\ \hline
Owl                         & 18.3k            & 307                    & 2.1k                & 38  \\ \hline
PraisonAI                   & 5.5k             & 263                    & 752                 & 28  \\ \hline
Ragflow                     & 66.9k            & 2811                   & 7.6k                & 377 \\ \hline
Sdk-python                  & 3.9k             & 270                    & 570                 & 72  \\ \hline
Smolagents                  & 24k              & 491                    & 2.2k                & 196 \\ \hline

\end{tabular}
\end{table}
\begin{table}[htbp]
\scriptsize
\caption{Domain categorization of the selected LLM-based multi-agent open-source projects}
\label{tab:MAS_domain_categorization}
\begin{tabular}{>{\raggedright}m{1.42cm}m{2.9cm}m{2.43cm}<{\centering}}
\hline
\textbf{Domain}         & \textbf{\#Definition}             & \textbf{\#Projects} \\ \hline
LLM-based Development Framework &
Open-source projects that provide frameworks for developing software systems by leveraging LLMs. &
MetaGPT, ChatDev, Agentscope, Langroid, Camel, Owl, AutoGen \\
\hline
Autonomous Agents and Multi-Agent Orchestration &
Open-source projects that focus on coordinating, managing, and enabling collaboration among multiple autonomous LLM-based agents. &
CrewAI, Agency-swarm, Ag2, Agno, PraisonAI \\
\hline
Agent SDKs and Toolkits &
Open-source software development kits that provide APIs and tools for building, deploying, and managing LLM-based agents. &
Adk-python, Openai-agents-python, Smolagents \\
\hline
Retrieval-Augmented LLM Applications &
Open-source projects that integrate information retrieval mechanisms with LLMs to enhance reasoning and response generation using external knowledge. &
Ragflow \\
\hline
Visual Orchestration and Workflow Automation &
Open-source platforms that enable visual, low-code, or automated orchestration of LLM workflows and agent pipelines. &
Flowise \\
\hline
AI Developer Tools &
Open-source tools designed to assist developers by integrating LLM-based capabilities into development environments and workflows. &
Continue \\
\hline
\end{tabular}
\end{table}

We prioritized projects with recent commit activity to ensure they reflect current MAS practices, and we manually verified whether each project explicitly described multi-agent collaboration rather than single-agent LLM usage. In addition, the exclusion criteria (E1) and (E2) were applied to remove educational, tutorial-oriented, or competition-based projects, as well as repositories that only reference multi-agent concepts without implementing actual multi-agent functionality. This selection process ensures that the final dataset is representative of established and actively maintained LLM-based MAS.

After applying the inclusion and exclusion criteria, we carefully examined the remaining repositories to ensure that they implemented LLM-based MAS and satisfied all selection requirements. Repositories that only partially met the criteria, lacked sufficient project information, or did not provide clear evidence of multi-agent functionality were excluded. Through this screening process, we identified a final set of 21 actively maintained open-source LLM-based MAS projects, which formed the basis of our subsequent analysis. Following the selection criteria, we ultimately included 21 open-source LLM-based MAS (see Table \ref{tbl:multi-agent-projects}) that met our criteria. The domains of the 21 selected LLM-based MAS projects are categorized into six categories (see Table \ref{tab:MAS_domain_categorization}). Then, the first author employed web scrapers to collect all the closed issues from these project repositories. In the end, we gathered a total of 22,848 closed issues. Each closed issue was recorded in an MS Excel file. We named each MS Excel file using the convention: ``\textsc{Project Name}'' + ``\textsc{\_}'' + ``\textsc{Issue Number}''.

\subsection{Data Filtering and Labeling}\label{subsec:DataFilteringandLabeling}
We collected closed issues from 21 open-source LLM-based MAS and observed that not all reported issues were relevant to MAS-related problems. For example, some practitioners used issue trackers for community interaction or general inquiries, such as requesting access to a project’s Discord channel \href{https://github.com/VRSEN/agency-swarm/issues/37}{\texttt{Agency-swarm} \#37}, without describing any technical problem in the MAS projects. We also identified issues that were purely single-agent in nature and did not involve multi-agent collaboration, despite being reported in MAS project repositories (e.g., \href{https://github.com/strands-agents/sdk-python/issues/1435}{\texttt{Sdk-python} \#1435}). In addition, we found that the issue labels were often too generic, and relying solely on existing labels could lead to missing relevant information. For example, the multi-agent \textit{ Orchestration Issue} reported in \href{https://github.com/microsoft/autogen/issues/6819}{\texttt{Autogen} \#6819} was assigned only the generic \texttt{Bug} label, making it difficult to identify through label-based filtering alone.

To address these challenges in filtering and labeling MAS-related issues, we adopted a two-stage mixed approach, combining automated and manual filtering methods to construct a high-quality dataset. This process focused on identifying closed issues that explicitly describe practitioner-reported problems related to MAS. Based on the pilot and formal filtering results in Section~\ref{subsubsec:PilotDataFiltering} and Section~\ref{subsubsec:FormalDataFiltering}, the final dataset was constructed and subsequently used for data extraction and analysis.

\subsubsection {Pilot Data Filtering}\label{subsubsec:PilotDataFiltering}
We conducted pilot data filtering by randomly selecting 100 closed issues from the collected dataset. The first and second authors independently labeled these issues to minimize potential personal bias. We observed that multi-agent issues were rare only 11 issues were related to real MAS issues, whereas non-multi-agent issues accounted for a relatively large proportion. Therefore, we adopted a mixed approach, combining automated and manual methods for data filtering, to ensure a high-quality dataset for our study. The pilot filtering results were compiled and recorded in MS Excel files \citep{replicationPackage}.
 
\emph{\textbf{Automated Filtering}}:
We applied a pilot automated filtering approach \citep{he2025llmjudge}, following the methodology of prior studies demonstrating the effectiveness of LLM-based evaluation and automated assessment in software engineering tasks \citep{zhou2025swe, crupi2025effectiveness}. We randomly selected 500 issues from the collected 22,848 closed issues and processed them with the \textit{DeepSeek-R1} model deployed locally via Ollama, with carefully crafted prompt instructions and iterative refinement to identify MAS-related issues~\citep{montes2025llmthematic}. Using this approach, we identified 110 MAS-related issues.

\emph{\textbf{Manual Filtering}}:
To ensure dataset quality, we did not fully rely on the LLM; therefore, manual filtering was also conducted. Two human coders, the first and second authors, were involved in this process. They labeled the same sample of 500 issues independently, which contained information related to problems encountered during the development or use of open-source LLM-based MAS. Inter-rater agreement between the two authors was measured using Cohen's Kappa coefficient \citep{cohen1960coefficient}, yielding a value of 0.82, which indicates a very high level of labeling consistency between the two authors. Any disagreements in the pilot labeling results were resolved through discussion involving the fourth author to reach a consensus. Finally, 102 out of the 500 closed issues were considered real multi-agent issues.

\emph{\textbf{Evaluation of Automated Filtering}}:
To evaluate the effectiveness and accuracy of the \textit{DeepSeek-R1} model, we compared the automated filtering results with those of the manual filtering. Out of the 500 sampled issues, manual filtering labeled 102 multi-agent-related issues, while automated filtering labeled 110 such issues. The comparison between manual and automated labels yielded 452 agreements, where both methods produced consistent labels (i.e., auto=0, manual=0 or auto=1, manual=1), and 48 disagreements, where the labels differed (i.e., auto=1, manual=0 or auto=0, manual=1). We measured the agreement between the automated filtering results and the human annotations. Using this measure, the automated filtering approach achieved an agreement rate of 90.4\%.  
To examine the 48 disagreements, we reviewed the cases where automated filtering labeled an issue as 0 but manual filtering labeled it as 1. In these cases, MAS concepts were not explicitly mentioned, and the relevance to MAS could only be determined from the issue context rather than explicit keywords. As a result, the LLM did not filter these issues as MAS-related issues. For example, \href{https://github.com/agentscope-ai/agentscope/issues/489}{\texttt{Agentscope} \#489} does not explicitly describe a multi-agent issue, but its relevance to MAS is only clear from the context, not from explicit keywords. Conversely, in cases where automated filtering labeled an issue as 1 but manual filtering labeled it as 0, the issues often contained MAS-related terminology or project components, leading the LLM to filter them as MAS-related issues. However, during manual filtering, they were determined not to clearly describe practitioner-reported MAS-related issues. For example, \href{https://github.com/MervinPraison/PraisonAI/issues/234}{\texttt{PraisonAI} \#234} reports a dependency installation error but uses the term \texttt{CrewAI}, which is a MAS framework. Based on this keyword, the LLM classified it as a MAS-related issue. Similarly, \href{https://github.com/ruvnet/ruflo/issues/571}{\texttt{Claude-flow} \#571} uses the phrase \texttt{Hive Mind}, which refers to MAS functionality, but it is actually a release announcement describing bug fixes and improvements rather than a practitioner-reported issue. 

\begin{table}[htbp]
\scriptsize
\caption{The source projects of the closed issues in the final data sample}
\label{tab:final_dataset_projects}
\begin{tabular}{>{\raggedright}m{2.4cm}m{1.05cm}<{\centering}m{1.32cm}<{\centering}m{1.42cm}<{\centering}}
\hline
\textbf{Project Name} & \textbf{\#Selected Issues} & \textbf{\#Contributors} & \textbf{\#Lines of Code} \\
\hline
Activepieces            & 6             & 336           & 1,358k \\ \hline
Adk-python              & 142           & 180           & 209k \\ \hline
Agency-swarm            & 27            & 21            & 79.1k \\ \hline
Agno                    & 126           & 343           & 390k \\ \hline
Ag2                     & 28            & 158           & 190k \\ \hline
Agentscope              & 23            & 23            & 71.1k \\ \hline
Autogen                 & 81            & 547           & 1,592.4k \\ \hline
Camel                   & 14            & 183           & 236k \\ \hline
ChatDev                 & 6             & 56            & 53.0k \\ \hline
Continue                & 34            & 421           & 661.7k \\ \hline
CrewAI                  & 77            & 278           & 466k \\ \hline
Claude-flow             & 70            & 11            & 1,489.7k \\ \hline
Flowise                 & 94            & 268           & 309.5k \\ \hline
Langroid                & 8             & 41            & 611.3k \\ \hline
MetaGPT                 & 44            & 113           & 70.1k \\ \hline
Openai-agents-python    & 22            & 182           & 108.1k \\ \hline
Owl                     & 7             & 38            & 14.2k \\ \hline
PraisonAI               & 31            & 28            & 512.7k \\ \hline
Ragflow                 & 8             & 377           & 414k \\ \hline
Sdk-python              & 39            & 72            & 59.3k \\ \hline
Smolagents              & 15            & 196           & 29.8k \\ \hline

\textbf{Total} & \textbf{944}   \\ \hline

\end{tabular}
\end{table}

\subsubsection{Formal Data Filtering } \label{subsubsec:FormalDataFiltering}
\begin{figure}[htbp]
    \centering
    \includegraphics[width=\columnwidth]{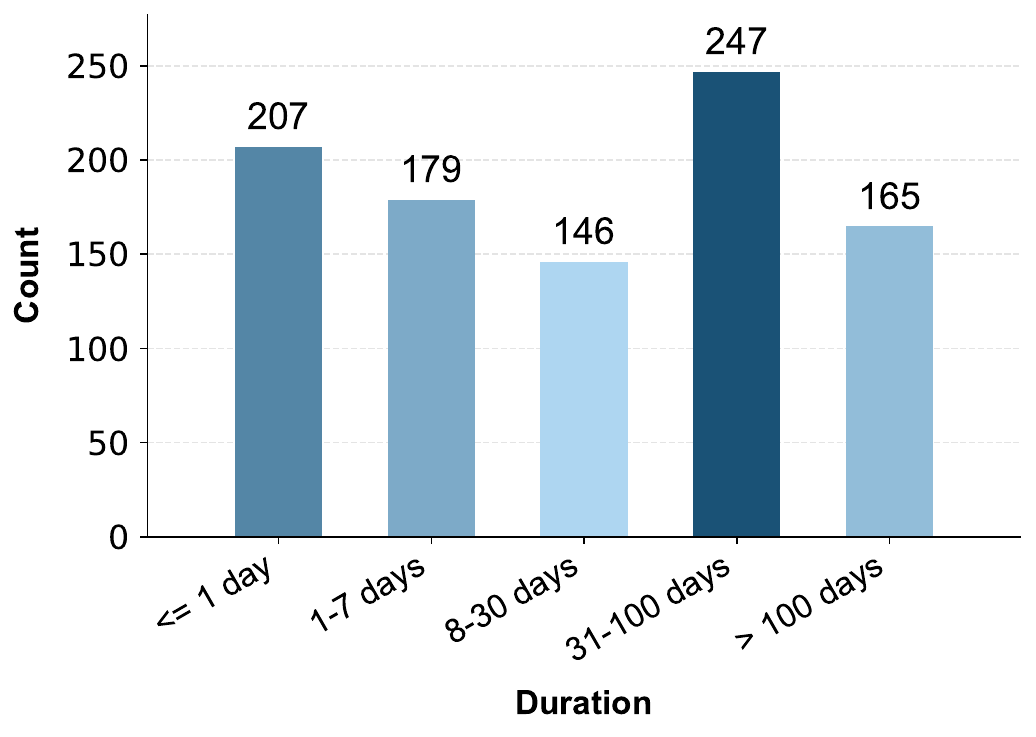}
    \caption{\textcolor{black}{The distribution of the duration of the closed issues in the final data sample}}
    \vspace{-0.75cm}
    \label{fig:SelectedDataSet}
\end{figure}

We conducted formal data filtering on the collected 22,848 closed issues. In this process, we excluded the issues not related to our study. We selected projects in which LLM-based MAS is a core component, and we kept those issues related to the problems that practitioners encountered during the development and use of open-source LLM-based MAS. As observed in the pilot study (see Section~\ref{subsubsec:PilotDataFiltering}), the proportion of real MAS issues among the collected 22,848 closed issues was relatively low. Therefore, we adopted a mixed approach, combining automated and manual methods to construct a high-quality dataset. Accordingly, we first applied automated filtering to all 22,848 collected closed issues using the \textit{DeepSeek-R1} model deployed locally via Ollama, with carefully crafted prompt instructions and iterative refinement~\citep{montes2025llmthematic}, to identify MAS-related issues~\citep{he2025llmjudge}. This automated filtering identified 2,335 multi-agent-related closed issues. To ensure dataset quality, we did not fully rely on the LLM; therefore, manual filtering was also conducted. The first author was involved in manually reviewing the results of the automated filtering to ensure data quality. Any disagreements in the formal data filtering results were resolved through discussion involving the fourth author to reach a consensus. Finally, 944 out of the 22,848 closed issues were considered real multi-agent issues. The source projects of the closed issues in the final dataset are presented in Table~\ref{tab:final_dataset_projects}. In addition, the distribution of the duration from issue opening to closing for the closed issues in the final dataset is shown in Figure~\ref{fig:SelectedDataSet}. The results of the formal data filtering were compiled and recorded in MS Excel files \citep{replicationPackage}.

\subsection {Data Extraction}\label{subsec:DataExtraction}
To answer the three RQs presented in Section~\ref{sec:goals}, we defined a set of data items for data extraction, as shown in Table~\ref{tab:data-items}. The data items D1--D3 are used to extract information about issues, underlying causes, and solutions that address these issues, corresponding to RQ1--RQ3, respectively. These data items are extracted from different parts of the collected closed issues, including the title, description, and discussion. To ensure the accuracy of the data extraction results and the validity of the subsequent data analysis, we decided to conduct the data extraction manually. In addition, the dataset size is relatively small, making manual data extraction feasible. The details of the pilot and formal data extraction procedures are presented in Section~\ref{subsubsec:pilotDataExtraction} and Section~\ref{subsubsec:FormalDataExtraction}, respectively.

\subsubsection {Pilot Data Extraction}\label{subsubsec:pilotDataExtraction}
We randomly selected 100 closed issues from our filtered dataset. The first author then independently performed pilot data extraction on these issues. The results indicate that all three data items (see Table~\ref{tab:data-items}) can be extracted from our dataset. Based on the pilot extraction results, we reached a consensus and defined the following rules for formal data extraction. (1) In principle, only one issue is extracted from each closed issue; if multiple issues can be extracted, we record the first issue discussed. (2) If there are multiple causes mentioned in a closed issue, we record all the causes of the issue extracted in the former step, which are identified by the reporter and the development team. (3) Similarly, if there are multiple solutions mentioned in a closed issue, we record all the solutions identified by the reporter and the development team that can address the issue extracted in the former step.

\begin{table}[htbp]
\scriptsize
\caption{Data items extracted and their corresponding RQs}
\label{tab:data-items}
\begin{tabular}{m{0.1cm}<{\centering}m{1.2cm}m{5.1cm}m{0.3cm}<{\centering}}
\hline{\#} & \textbf{Data Item} & \textbf{Description} & \textbf{RQ} \\
\hline
D1 & Issue &
The main issue discussed in a GitHub issue report. In principle, only one issue is extracted from each closed issue. &
RQ1 \\
\hline
D2 & Cause &
The cause(s) of the issue are identified. Due to the complexity of some issues, multiple causes may exist. &
RQ2 \\
\hline
D3 & Solution &
The solution(s) that address the issue are provided. Some issues may have multiple solutions. &
RQ3 \\
\hline
\end{tabular}
\end{table}

\subsubsection {Formal Data Extraction}\label{subsubsec:FormalDataExtraction}
Following the pilot phase, the first author independently performed the formal data extraction based on the items defined in Table~\ref{tab:data-items}. Any ambiguities encountered during the data extraction process were resolved through discussions with the fourth author to ensure consistency. After completing the data extraction, the second and fourth authors reviewed the results and discussed them with the first author to resolve inconsistencies and reach a consensus. Both authors further conducted multiple rounds of review and refinement to ensure the accuracy of the final dataset. The final results were compiled and recorded in an MS Excel file~\citep{replicationPackage}.

During the formal data extraction, we identified 16 closed issues that contained multiple problems within a single GitHub issue report (e.g., \href{https://github.com/crewAIInc/crewAI/issues/3918}{\texttt{CrewAI \#3918}}). In accordance with our predefined criteria in Section~\ref{subsubsec:pilotDataExtraction}, we included only the first reported problem along with its corresponding causes and solutions. We also observed that not all closed issues contained clearly extractable information (i.e., causes or solutions). In some cases, the discussion was too vague to identify specific causes or solutions, while in others, issues were automatically closed due to inactivity, low priority, or high complexity, making it difficult to extract information.

\subsection {Data Analysis}\label{subsec:DataAnalysis}
After completing the data extraction, we performed data analysis to address the three RQs formulated in Section~\ref{sec:goals}. We applied the Open Coding and Constant Comparison methods, which are widely used in Grounded Theory for qualitative analysis in software engineering research~\citep{stol2016grounded}. Open Coding represents the initial stage of analysis, where researchers systematically examine the text to identify key concepts and categories, assigning descriptive codes to each element. Constant Comparison follows as a refinement stage, in which researchers iteratively compare new data with previously coded data to adjust and integrate categories based on their similarities and differences.

The steps in our data analysis are as follows: (1) The first author meticulously read through the text of each closed issue and assigned descriptive codes that succinctly summarize the core themes according to the three RQs (see Table~\ref{tab:data-items}). For example, \href{https://github.com/huggingface/smolagents/issues/1353}{\texttt{Smolagents} \#1353} was coded as \texttt{``Agent Tool Registration Failure''}, which indicates an error caused by improper agent integration. (2) The first author compared commonalities and differences among codes. Following a bottom-up approach, identical or similar codes were grouped into higher-level types and categories. For example, the code from \href{https://github.com/huggingface/smolagents/issues/1353}{\texttt{Smolagents} \#1353}, along with similar instances, constituted a \textit{Tool Invocation Error} type under the broader \textit{Tool Issue} category. (3) When uncertainties about code descriptions arose, the first author discussed them with the second and fourth authors to reach a consensus. Due to the iterative nature of constant comparison, the initial results underwent several modifications. (4) The fourth author reviewed and verified the initial analysis. Subsequently, all other authors reviewed the final results and made further refinements to improve the accuracy and consistency of the analysis. Any disagreements were resolved through discussion; after resolving 29 conflicts in issue categories, 24 in cause categories, and 22 in solution categories, we obtained the final data analysis results.

\section{Results and Interpretation}\label{sec:ResultsandInterpretation}
\begin{figure*}[h]
    \centering
    \includegraphics[width=\linewidth]{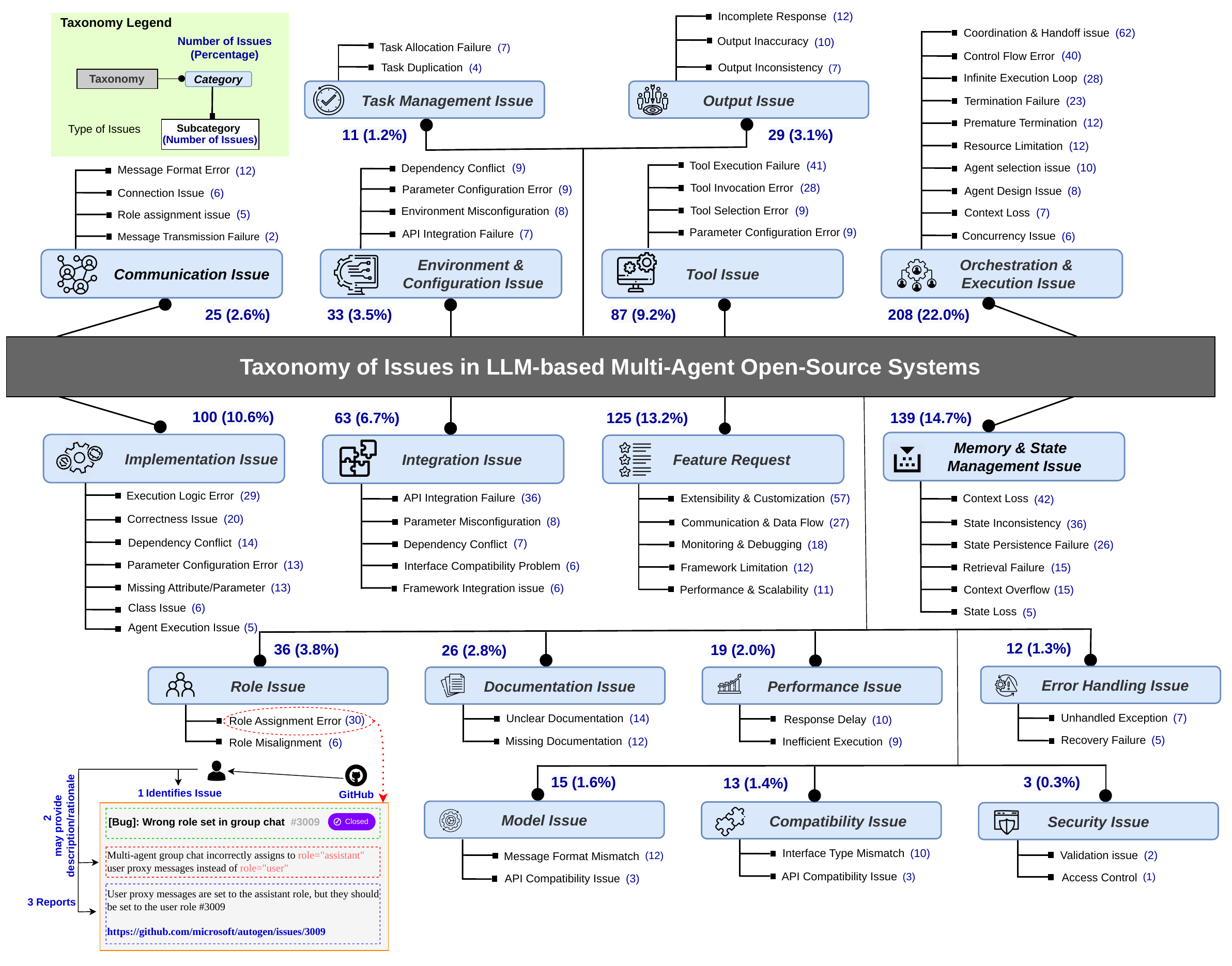}
    \caption{Taxonomy of issues of open-source LLM-based MAS}
    \label{fig:taxonomy_issues}
\end{figure*}

We present the findings of the three RQs and discuss their implications. We introduce a two-stage classification for the Issue, Cause, and Solution taxonomies, where an Issue refers to a problem reported by practitioners in LLM-based MAS, a Cause represents the underlying reason for the problem, and a Solution describes the approach used to resolve it. The taxonomies consist of categories and types, and brief descriptions of each type are provided under the corresponding category. To demonstrate the relationships among issues, causes, and solutions, we present mappings between Issue and Cause categories, as well as between Issue and Solution categories. It should be noted that only closed issues for which a solution has been extracted are included in the issue solution mapping. Similarly, only closed issues for which a cause has been extracted are included in the issue cause mapping. The taxonomy of issues in the final dataset is shown in Figure~\ref{fig:taxonomy_issues}, while the distribution of their closure durations is presented in Figure~\ref{fig:SelectedDataSet}. To further interpret these findings, we identify key patterns in the results and examine their implications for issues, causes, and solutions in LLM-based MAS.

\subsection {Category of Issues (RQ1)} \label{subsec:CategoryofIssues}
Figure~\ref{fig:taxonomy_issues} shows the taxonomy of the closed issues extracted from the dataset. The results show that \emph{Orchestration \& Execution Issue} (22.0\%) is the most frequent category reported by practitioners in open-source LLM-based MAS projects, followed by \emph{Memory \& State Management Issue} (14.7\%) and \emph{Feature Request} (13.2\%). \emph{Implementation Issue} (10.6\%) also represents a notable portion of the dataset. The remaining categories each account for less than 10\%, including \emph{Tool Issue} (9.2\%), \emph{Integration Issue} (6.7\%), \emph{Role Issue} (3.8\%), \emph{Environment \& Configuration Issue} (3.5\%), \emph{Output Issue} (3.1\%), \emph{Documentation Issue} (2.8\%), \emph{Communication Issue} (2.6\%), \emph{Performance Issue} (2.0\%), \emph{Model Issue} (1.6\%), \emph{Compatibility Issue} (1.4\%), \emph{Error Handling Issue} (1.3\%), \emph{Task Management Issue} (1.2\%), and \emph{Security Issue} (0.3\%).

\subsubsection{Orchestration \& Execution Issue} (22.0\%)\label{subsec:OrchestrationExecutionIssues}
\emph{Orchestration \& Execution Issue} refers to problems arising from agent orchestration and LLM-based execution control during runtime in the development and use of open-source LLM-based MAS. This category comprises the following ten issue types.

\begin{itemize}
    \item \textbf{\textsc{Coordination \& Handoff Issue}} refers to issues in interaction and task transfer between agents in LLM-based MAS. These issues usually arise when agents fail to properly coordinate or hand off tasks, leading to delays, duplication, unexpected outputs, or incomplete execution. For example, a practitioner reported that ``\textit{sub-agent could not correctly transfer to root agent and call the MCP tool}'' (Adk-python \#1039).
    
    \item \textbf{\textsc{Control Flow Error}} refers to issues where the execution order between agents or tasks does not behave as expected. These issues include incorrect sequencing, unexpected loop behavior, or disrupted execution flow, which may lead to incorrect outputs, execution failure, or early termination. For example, a practitioner reported that ``\textit{SequentialAgent doesn't execute sub-agents after the first one}'' (Adk-python \#573).
    
    \item \textbf{\textsc{Infinite Execution Loop}} refers to issues where agents become trapped in continuous execution without a valid termination condition. These issues arise from missing termination logic, weak control, or poor coordination, leading to resource waste and incomplete workflows. For example, a practitioner reported that ``\textit{agentic parallelization with different agents/tasks: tasks repeat in a loop, aggregator never runs, process never completes}'' (PraisonAI \#864).
    
    \item \textbf{\textsc{Termination Failure}} refers to issues where execution continues even after a valid termination condition is met. These issues result from incorrect termination signaling or inconsistent stopping conditions, causing workflows to run longer than necessary. For example, a practitioner reported that ``\textit{termination on the receiving agent doesn't terminate the program}'' (Autogen \#1919).
    
    \item \textbf{\textsc{Premature Termination}} refers to issues where execution stops before all agents complete their tasks. These issues arise from incorrect orchestration, early stopping, or faulty propagation, resulting in partial workflow execution. For example, a practitioner reported that ``\textit{multi-agent chain stops at first agent without executing others in v2.2.7}'' (Flowise \#4321).
    
    \item \textbf{\textsc{Resource Limitation}} refers to issues where LLM-based MAS experience resource conflicts, shared tool contention, or time constraints. These issues result in execution hangs, failures, or early termination. For example, a practitioner reported that ``\textit{Claude-Flow swarm command consistently times out due to resource conflicts between the swarm execution and active Claude Code sessions}'' (Claude-flow \#59).
    
    \item \textbf{\textsc{Agent Selection Issue}} refers to issues where the routing or delegation mechanism fails to identify and assign the most appropriate specialized agent for a given task. These issues lead to incorrect agent selection or failed agent delegation decisions, resulting in incorrect or failed task execution. For example, a practitioner reported that ``\textit{manager llm doesn't seem to be working properly and does not seem to delegate the query to other agents}'' (CrewAI \#1503).
    
    \item \textbf{\textsc{Agent Design Issue}} refers to issues where agent architectural design, data flow, or inter-agent communication is difficult to define or implement in LLM-based MAS. These issues lead to challenges in chaining agents, handling structured data, or maintaining consistency across agent boundaries, resulting in orchestration errors or workflow breakdowns. For example, a practitioner reported that ``\textit{passing information between 2 agents is cumbersome}'' (Agno \#2391).
    
    \item \textbf{\textsc{Context Loss}} refers to issues where runtime contextual information (e.g., domain, task constraints, or user instructions) is not properly propagated between agents during execution. Agents ignore or revert to default values instead of following the intended context, leading to incorrect or unexpected behavior. For example, a practitioner reported that ``\textit{agents failed to consistently follow the specified domain context during multi-agent execution, repeatedly using the wrong domain example.com}'' (PraisonAI \#289).
    
    \item \textbf{\textsc{Concurrency Issue}} refers to issues where shared state is not properly managed when multiple agents execute in parallel. These issues arise when actions from different agents are processed in an unexpected order, potentially leading to inconsistent or incomplete context, runtime errors, or unstable behavior. For example, a practitioner reported that ``\textit{when running multiple agents in parallel, events coming from different agents can interleave}'' (Adk-python \#2404).
\end{itemize}

\textbf{\emph{Interpretation}}: The high proportion of \emph{Orchestration \& Execution Issue} indicates that runtime coordination is the main challenge in LLM-based MAS. \emph{Coordination \& Handoff Issue} (62) and \emph{Control Flow Errors} (40) are the most frequent types of issues in this category. \emph{Coordination \& Handoff Issue} reflect failures in task transfer between agents, while \emph{Control Flow Error} indicate incorrect or unstable execution ordering. Overall, most issues arise from weak coordination between agents rather than limitations in individual agents. This indicates a gap between theoretical multi-agent design and practical execution reliability. The findings highlight orchestration as a key bottleneck and the need for better execution control and coordination mechanisms in future LLM-based MAS.

\subsubsection{Memory \& State Management Issue} (14.7\%)\label{subsec:MemoryStateManagementIssue}
\emph{Memory \& State Management Issue} refers to issues in storing, updating, and propagating memory, state, or intermediate information across agents during execution. Practitioners encounter these challenges during the development and use of LLM-based multi-agent OSS. This category encompasses the following six types.

\begin{itemize}
    \item \textbf{\textsc{Context Loss}} refers to issues where persistent state or shared memory is not properly maintained or retrieved across agents during execution. Contextual information stored at higher levels is not consistently propagated or is reinitialized instead of being reused, causing agents to lose access to previously established states and leading to inconsistent or incorrect behavior across execution steps. For example, a practitioner reported that ``\textit{team shared context is lost whenever moving on to subsequent agent}'' (Agno \#3254).
    
    \item \textbf{\textsc{State Inconsistency}} refers to issues where shared state or intermediate outputs are not consistently visible, synchronized, or aligned across agents during execution. State updates are not reliably propagated, or different agents access different versions of the same state due to timing or synchronization issues, causing agents to operate on inconsistent state views, leading to unreliable or divergent execution behavior. For example, a practitioner reported that ``\textit{the second agent is often unable to see items stored in the state of the previous agent}'' (Adk-python \#1119).
    
    \item \textbf{\textsc{State Persistence Failure}} refers to issues where stored state or session-level information is not correctly initialized, preserved, or made accessible across agents during execution. State is not properly bound to session identifiers, or stored information becomes unavailable during runtime despite being explicitly defined, preventing agents from retrieving previously stored session state and leading to incomplete or incorrect execution. For example, a practitioner reported that ``\textit{team\_session\_state not initialized when session\_id is provided}'' (Agno \#3799).
    
    \item \textbf{\textsc{Retrieval Failure}} refers to issues where previously stored state, memory entries, or checkpoints cannot be reliably accessed during execution, even if they exist in the system. Agents fail to retrieve the required previous state, leading to incomplete workflow progression or reduced system reliability. For example, a practitioner reported that ``\textit{in a sequential agent workflow, the system randomly fails to retrieve stored checkpoints}'' (Flowise \#3859).
    
    \item \textbf{\textsc{Context Overflow}} refers to issues where accumulated contextual information exceeds the effective input or memory capacity of the system during multi-agent execution. Multiple agents continuously contribute information without filtering or compression, making the input context excessively large and leading to degraded performance or loss of relevant signals during reasoning. For example, a practitioner reported that ``\textit{a single agent needs to store information from multiple agents, which can lead to excessively long token lengths in the input for large models}'' (Agentscope \#637).
    
    \item \textbf{\textsc{State Loss}} refers to issues where parts of the generated or updated state are not retained in the final execution outcome. In multi-agent or parallel workflows, intermediate state updates are overwritten, dropped, or not properly merged during aggregation, causing some agent contributions to be missing from the final system state and resulting in incomplete or partial execution results. For example, a practitioner reported that ``\textit{state updates from parallel agent calls were not correctly aggregated, causing some agent contributions to be missing from the final session state}'' (Adk-python \#1153).
\end{itemize}

\textbf{\emph{Interpretation}}: The high proportion of \emph{Memory \& State Management Issue} indicates that state handling is a critical challenge in LLM-based MAS. \textit{Context Loss} (42) and \textit{State Inconsistency} (36) are the most frequently reported issues, showing a consistent pattern of unreliable state propagation across agents. There is a clear gap between the designed memory abstractions and their actual behavior in execution. This highlights the need for reliable state management and memory propagation mechanisms in future MAS.

\subsubsection {Feature Request} (13.2\%) \label{subsec:FeatureRequest}
\emph{Feature Request} refers to practitioner-driven requests for new capabilities or enhancements in LLM-based MAS. These requests typically highlight gaps between existing system capabilities and real-world application needs. This category encompasses the following five types.

\begin{itemize}
    \item \textbf{\textsc{Extensibility \& Customization}} refers to requests for extending system flexibility to support custom configurations, models, or agent-specific behaviors in LLM-based MAS. For example, a practitioner requested ``\ textit {Is it possible to define multiple (local) LLMs in the config.yaml for different agents?}'' (MetaGPT \#1266).
    
    \item \textbf{\textsc{Communication \& Data Flow}} refers to requests for improving how information is transmitted, streamed, or exposed across agents and system components in LLM-based MAS. These requests typically arise when intermediate outputs or interactions are not visible or are delayed during execution. For example, a practitioner requested ``\textit{a way to output or stream AgentFlows like Chat inside Flowise app}'' (Flowise \#2838).
    
    \item \textbf{\textsc{Monitoring \& Debugging}} refers to requests for improving visibility into agent behavior, execution steps, and system-level activities in LLM-based MAS. These requests typically arise when practitioners cannot observe how agents make decisions or interact with tools and data during execution. For example, a practitioner requested ``\textit{observability of agents' actions taken, tools used, and data accessed}'' (AutoGen \#2423).
    
    \item \textbf{\textsc{Framework Limitation}} refers to requests in multi-agent frameworks where certain capabilities or design requirements are not directly supported by the existing system design or API. These requests typically arise when the framework only supports simple sequential or instruction-based workflows, thereby limiting more structured orchestration designs. For example, a practitioner requested support for advanced orchestration patterns: ``\textit{The agent team API cannot handle complicated workflows or hierarchical multi-agent patterns}'' (Agno \#2209).
    
    \item \textbf{\textsc{Performance \& Scalability}} refers to requests aimed at improving execution speed, parallelism, and system throughput in multi-agent frameworks. These requests typically arise when current systems do not efficiently support large-scale agent execution or real-time control over running processes. For example, a practitioner proposed ``\textit{parallel agent spawning to improve execution speed and support real-time query control in multi-agent workflows}'' (Claude-flow \#781).
\end{itemize}

\textbf{\emph{Interpretation}}: The \emph{Feature Request} category shows that practitioners mainly focus on extending the capabilities of existing multi-agent frameworks. A clear pattern emerges in \emph{Extensibility \& Customization} (57) and \emph{Communication \& Data Flow} (27), where users request more flexible configurations and support for advanced design capabilities not directly available in existing frameworks. Overall, these findings indicate that current MAS frameworks are functional but still limited in flexibility, leading practitioners to request more adaptable and extensible system designs.

\subsubsection {Implementation Issue} (10.6\%) \label{subsec:ImplementationIssue}
\emph{Implementation Issue} refers to issues arising from incorrect or incomplete implementation in LLM-based MAS during development or runtime. This category encompasses the following seven types.

\begin{itemize}
    \item \textbf{\textsc{Execution Logic Error}} refers to issues arising from incorrect or unintended execution behavior caused by flaws in implementation logic during runtime in LLM-based MAS. Function calls, control flow, or internal execution mechanisms do not behave as intended, leading to runtime errors or incorrect execution paths despite a correct intended design. For example, a practitioner reported that when an agent is copied, method references become mismatched: ``\textit{the copied version of the BoardAgent causes a KeyError when accessing hook\_lists[self.process\_all\_messages]}'' (AutoGen \#1706).
    
    \item \textbf{\textsc{Correctness Issue}} refers to issues where the system produces factually incorrect, logically inconsistent, or semantically invalid results during execution in LLM-based MAS. Agent reasoning or intermediate computations deviate from expected functional behavior despite successful execution, resulting in wrong answers, miscalculations, or flawed decisions even when the output format is correct. For example, a practitioner reported that the system returns incorrect stock prices in a multi-agent setup: ``\textit{multi-agent framework not working as expected, issue with finance lib as well internally}'' (Agno \#1700).
    
    \item \textbf{\textsc{Dependency Conflict}} refers to issues where different framework or library versions lead to incompatible behaviors in LLM-based MAS. These issues arise when updates in dependencies modify core execution logic or agent interaction mechanisms, causing regressions in previously working workflows or system failures. For example, a practitioner reported that a version update breaks the group chat speaker selection logic: ``\textit{Autogen Studio GroupChatWorkflow not working on pyautogen==0.2.13 but working on pyautogen==0.2.10}'' (AutoGen \#1693).
    
    \item \textbf{\textsc{Parameter Configuration Error}} refers to issues where incorrect or inconsistent parameter passing, or misconfiguration in setup, leads to unexpected behavior in LLM-based MAS during execution. These issues typically arise when input parameters are not properly propagated across agents, tools, or execution modes under different runtime settings, causing the system to execute successfully but fail to correctly utilize required inputs, resulting in incomplete behavior. For example, a practitioner reported that enabling streaming causes the files argument to be lost during execution: ``\textit{Files parameter not passed to tools when using team streaming}'' (Agno \#5121).
    
    \item \textbf{\textsc{Missing Attribute/Parameter}} refers to implementation issues where required attributes or parameters are missing or incorrectly defined in data models or execution objects in LLM-based MAS. Framework components evolve, but dependent modules or logging/processing layers are not updated accordingly, leading to attribute mismatches during runtime. For example, a practitioner reported that ``\textit{AttributeError: `Message' object has no attribute `params'}'' (Adk-python \#2930).
    
    \item \textbf{\textsc{Class Issue}} refers to implementation issues where class types are not properly registered, recognized, or handled during runtime in LLM-based MAS. Custom or extended class types (including structured outputs) may not be supported by existing runtime components, leading to execution errors. For example, a practitioner reported that custom structured message classes are not recognized by the group chat runtime: ``\textit{agent's StructuredMessage class is not registered in GroupChat}'' (AutoGen \#6983).
    
    \item \textbf{\textsc{Agent Execution Issue}} refers to issues where agents do not follow the intended execution strategy (e.g., parallelism, scheduling, or control flow) during runtime. The orchestration layer may enforce unintended sequential processing, leading to agents executing in a different order than expected. For example, a practitioner reported that ``\textit{graph nodes execute sequentially instead of parallel}'' (Sdk-python \#478).
\end{itemize}

\textbf{\emph{Interpretation}}: The substantial proportion of \emph{Implementation Issue} indicates that correct runtime behavior is a critical challenge in LLM-based MAS. \emph{Execution Logic Error} (29) and \emph{Correctness Issue} (20) highlight a consistent pattern of faulty execution flow and incorrect system behavior. There is a clear gap between intended system design and its implementation, as implementation flaws lead to incorrect execution behavior, dependency conflicts, parameter misconfigurations, and other runtime issues in LLM-based MAS.

\subsubsection {Tool Issue} (9.2\%) \label{subsec:ToolIssue}
\emph{Tool Issue} refers to issues arising from incorrect integration, configuration, or usage of external tools in LLM-based MAS. This category encompasses the following four types.

\begin{itemize}
    \item \textbf{\textsc{Tool Execution Failure}} refers to issues that arise when tools fail during the runtime execution of LLM-based MAS. Tool invocation mechanisms or execution environments are improperly handled, leading to crashes or incomplete execution despite correct setup. For example, a practitioner reported that improper async handling causes a runtime error when invoking tools in a group chat: ``\textit{GroupChat cannot run with MCP tool call (async await issue)}'' (Ag2 \#1763).
    
    \item \textbf{\textsc{Tool Invocation Error}} refers to issues where tools are incorrectly invoked during execution in LLM-based MAS. Function names, mappings, or invocation logic are improperly configured, leading to calls to invalid or non-existent tools and failure to execute the intended functions. For example, a practitioner reported that missing function mappings cause incorrect tool selection: ``\textit{AutoGen tries to call non-existent functions that are the names of agents}'' (AutoGen \#781).
    
    \item \textbf{\textsc{Tool Selection Error}} refers to issues where agents improperly select or access tools that are not intended for them in LLM-based MAS. These issues typically arise when tool access boundaries are not properly defined in the multi-agent group configuration or when tools assigned to one agent become available to other agents, allowing these agents to use tools beyond their intended scope. For example, a practitioner reported that an agent uses tools registered for another agent due to shared group context: ``\textit{Why agent can use tools from another agent in autogen}'' (AutoGen \#4399).
    
    \item \textbf{\textsc{Parameter Configuration Error}} refers to issues arising from incorrect or incompatible parameter definitions in tool schemas within LLM-based MAS. The selected tool is correct, but the provided inputs do not align with the expected tool schema in terms of structure, type, or format, leading to validation failures before tool execution. For example, a practitioner reported that ``\textit{the manager agent passes dictionary values where the delegation tool schema expects strings}'' (CrewAI \#2606).
\end{itemize}

\textbf{\emph{Interpretation}}: The \emph{Tool Issue} category highlights key challenges in LLM-based MAS related to tool usage during execution. The most common issues include \emph{Tool Execution Failure} (41) and \emph{Tool Invocation Error} (28). These findings indicate that issues primarily arise from runtime breakdowns and incorrect tool selection in multi-agent settings, reflecting a gap between intended tool behavior and actual system execution.

\subsubsection {Integration Issue}(6.7\%)\label{subsec:IntegrationIssue}
\emph{Integration Issues} refer to issues related to the interaction and compatibility between different components in LLM-based MAS that practitioners encounter while developing and using these MAS. This category encompasses the following five types.

\begin{itemize}
    \item \textbf{\textsc{API Integration Failure}} refers to issues where external APIs or service-based models fail to integrate properly within LLM-based MAS. Integration issues between certain APIs or live services during execution result in runtime errors, thereby limiting communication between agents and external services. For example, a practitioner reported that live API constraints prevent the proper integration of sub-agent functionality: ``\textit{Live Event, Live API stuffs cannot use sub-agent}'' (Adk-python \#491).
    
    \item \textbf{\textsc{Parameter Misconfiguration}} refers to issues where parameter definitions used during the integration of external models or APIs do not align with the expected schema in LLM-based MAS. The selected model or service is correct, but the parameters passed across system components are improperly structured, missing required fields, or incompatible with the API specification. For example, a practitioner reported that misconfigured tool parameters cause schema validation errors: ``\textit{Gemini models in agent mode giving error 400 invalid argument}'' (Continue \#6121).
    
    \item \textbf{\textsc{Dependency Conflict}} refers to issues in which incompatible or mismatched library dependencies hinder the successful integration of components in LLM-based MAS. Different system environments rely on conflicting package versions or unsupported dependency combinations. For example, a practitioner reported that dependency incompatibilities lead to integration failures during deployment due to serialization errors: ``\textit{GroupChatWorkflow breaks after upgrading from pyautogen==0.2.10 to 0.2.13 due to incompatible dependency behavior}'' (Autogen \#1693).
    
    \item \textbf{\textsc{Interface Compatibility Problem}} refers to issues arising from incompatibility between different interfaces or execution environments in LLM-based MAS. The same agentic workflow behaves correctly in a local or direct execution environment but fails when integrated with external interfaces such as web sockets, IO streams, or service-based communication layers. For example, a practitioner reported that a group chat system fails to execute function calls when running through a WebSocket integration, despite working correctly in a direct environment: ``\textit{Group-chat will not call function when using IOStream}'' (AutoGen \#2824).
    
    \item \textbf{\textsc{Framework Integration Issue}} refers to issues where external multi-agent frameworks or libraries fail to integrate correctly within LLM-based MAS. Developers combine different agent frameworks or extend reference implementations, but runtime behavior breaks due to missing framework-specific configurations, incompatible abstractions, or incorrect usage of required components. For example, a practitioner reported that a SemanticKernel-based multi-agent setup fails when adapting a reference group-chat implementation, resulting in runtime exceptions: ``\textit{Write sample based on SemanticKernel with content copied from Example04\_Dynamic\_GroupChat\_Coding\_Task, can't work}'' (AutoGen \#2688).
\end{itemize}

\textbf{\emph{Interpretation}}: The \emph{Integration Issue} category highlights key challenges in LLM-based MAS related to integrating external APIs, frameworks, and system components. The most common issues include \emph{API Integration Failure} (36) and \emph{Parameter Misconfiguration} (8). Overall, these findings indicate that even when components are present and accessible, failures mainly arise from incorrect integration and mismatched assumptions across the system. Overall, these findings indicate that integration issues stem primarily from incompatibilities and configuration mismatches between external components and the MAS.

\subsubsection{Role Issue} (3.8\%) \label{subsec:RoleIssue}
\emph{Role Issue} refers to issues related to incorrect or unclear role assignment in LLM-based MAS, where agents fail to consistently follow their intended roles during execution. This category encompasses the following two types.

\begin{itemize}
    \item \textbf{\textsc{Role Assignment Error}} refers to issues where LLM-based MAS fail to correctly assign or evaluate roles during execution. The role assignment mechanism or routing logic does not produce valid or sufficient role decisions, leading to execution failure or inability to proceed with multi-agent coordination. For example, a practitioner reported that ``\textit{role assignment fails when the system receives insufficient compatibility scores for the user role, preventing the multi-agent process from proceeding}'' (CAMEL \#419).
    
    \item \textbf{\textsc{Role Misalignment}} refers to issues where LLM-based MAS store incorrect role labels in logs or chat history, even when the execution flow is correct. Message handling or logging mechanisms record wrong role tags (e.g., user messages recorded as assistant), leading to inconsistencies between actual agent behavior and the recorded system state. For example, a practitioner reported that messages from a user proxy agent are incorrectly recorded as assistant in the chat history: ``\textit{Wrong role set in group chat}'' (AutoGen \#3009).
    \end{itemize}
    
\textbf{\emph{Interpretation}}: The \emph{Role Issue} category highlights key challenges in LLM-based MAS related to correct role assignment and role labeling. These issues comprise \emph{Role Assignment Error} (30) and \emph{Role Misalignment} (6). Overall, role-related problems affect both role behavior during execution and role recording in logs, leading to inconsistencies between actual agent behavior and stored role information.

\subsubsection {Environment \& Configuration Issue} (3.5\%) \label{subsec:EnvironmentConfigurationIssue}
\emph{Environment \& Configuration Issue} refers to issues arising from incorrect setup, dependency management, or runtime environment configuration in LLM-based MAS, where agents fail to initialize or execute correctly. This category encompasses the following four types.

\begin{itemize}
    \item \textbf{\textsc{Dependency Conflict}} refers to issues where incompatible library or package versions in the execution environment prevent proper initialization or execution of LLM-based MAS. These runtime environment mismatches lead to runtime failures such as missing attributes or import errors, even when the system logic is correctly implemented. For example, a practitioner reported that a version mismatch between \texttt{autogen\_studio} and \texttt{autogen\_agentchat} causes a missing attribute error during team initialization: ``\textit{Type object team has no attribute load\_component with AutoGen Studio v0.4.0.4}'' (AutoGen \#5319).
    
    \item \textbf{\textsc{Parameter Configuration Error}} refers to issues where parameter configuration is incorrectly structured or not properly mapped to target agents or nodes in the current execution environment in LLM-based MAS. Runtime parameter overrides fail when schema mismatches, invalid formatting, or incorrect node identifiers, leading to execution errors or ignored settings despite correct system logic. For example, a practitioner reported that structured override configurations for sequential agents and LLM nodes cause runtime errors: ``\textit{Can't override data for sequential agents}'' (Flowise \#3240).
    
    \item \textbf{\textsc{Environment Misconfiguration}} refers to issues where LLM-based MAS fail when incorrect or incompatible deployment environment settings, such as networking rules, service exposure, or runtime constraints in cloud or containerized setups. These issues prevent agents from accessing external services or internal APIs, leading to connection failures despite correct system logic. For example, a practitioner reported that agent creation fails because of Kubernetes networking restrictions that block access to external AI endpoints: ``\textit{AI agents fail to create due to Kubernetes networking constraints}'' (Activepieces \#9038).
    
    \item \textbf{\textsc{API Integration Failure}} refers to issues where external APIs or tool interfaces fail to operate correctly due to environmental configuration constraints in LLM-based MAS. Incorrect lifecycle management or missing integration handling between the agent framework and external API services prevents agents from accessing API-based tools despite correct system logic. For example, a practitioner reported that MCP tool connections fail inside a FastAPI-mounted application because of improper lifecycle handling of the API interface: ``\textit{mounting AgentOS app on existing FastAPI prevents MCP tools connection}'' (Agno \#4958).
\end{itemize}

\textbf{\emph{Interpretation}}: The \emph{Environment \& Configuration Issue} category highlights key challenges in LLM-based MAS related to deployment setup, dependency management, and external service integration. The most common issues include \emph{Dependency Conflict} (9) and \emph{Parameter Configuration Error} (9). Overall, these findings indicate that environment \& configuration issues can prevent agents from initializing, communicating, or executing correctly within LLM-based MAS.

\subsubsection {Output Issue} (3.1\%) \label{subsec:OutputIssue}
\emph{Output Issue} refers to issues related to incorrect, incomplete, or malformed outputs generated by LLM-based MAS during execution that practitioners encounter while developing and using these systems. This category encompasses the following three types.

\begin{itemize}
    \item \textbf{\textsc{Incomplete Response}} refers to issues where LLM-based MAS return only a subset of the expected output, omitting intermediate results, agent-level responses, or other information from the execution process. This reduces transparency and makes it difficult to understand how the final answer was produced. For example, a practitioner reported that only the final conclusion is returned through the API while intermediate worker outputs are hidden: ``\textit{API of Agentflows not displaying the answers of each worker}'' (Flowise \#2706).
   
    \item \textbf{\textsc{Output Inaccuracy}} refers to issues where LLM-based MAS generate outputs containing unintended formatting artifacts or structural tokens in the final response, such as extra symbols, wrappers, or formatting noise not intended in the output format. This makes the output unsuitable for direct use or downstream processing. For example, a practitioner reported that ``\textit{triple backticks (```) are appended unexpectedly at the end of crew output responses}'' (CrewAI \#2105).
    
    \item \textbf{\textsc{Output Inconsistency}} refers to issues where LLM-based MAS produce inconsistent or unstable outputs under the same input conditions, even when the prompt, model, and configuration remain unchanged. These inconsistencies may arise from variations in multi-agent coordination, task decomposition, or sequential execution, leading to differences in the generated outputs. For example, a practitioner reported that ``\textit{Crew fails to deliver the same output as ChatGPT given the same prompt and model}'' (CrewAI \#1024).
\end{itemize}

\textbf{\emph{Interpretation}}: The \emph{Output Issue} category highlights key challenges in LLM-based MAS related to the completeness, correctness, and reliability of generated outputs. The most common issues include \emph{Incomplete Response} (12) and \emph{Output Inaccuracy} (10). Overall, these findings indicate that incomplete, malformed, and inconsistent outputs can reduce the usability and reliability of responses generated by LLM-based MAS.

\subsubsection {Documentation Issue}(2.8\%)\label{subsec:DocumentationIssue}
\emph{Documentation Issue} refers to issues related to missing, unclear, or incorrect documentation in LLM-based MAS. This category encompasses the following two types.

\begin{itemize}
    \item \textbf{\textsc{Unclear Documentation}} refers to issues where system documentation is inconsistent, ambiguous, or incorrect, leading to confusion during development and usage of LLM-based MAS. Documentation-related information such as agent and tool descriptions, usage instructions, or multi-agent workflows may not align with the actual implementation in LLM-based MAS. For example, a practitioner reported that documentation references outdated tool prefixes, causing naming confusion: ``\textit{confusing MCP Tool References - mcp\_ruv-swarm vs mcp\_claude-flow}'' (Claude-flow \#160).

    \item \textbf{\textsc{Missing Documentation}} refers to issues where required information, commands, or usage details are not properly documented in LLM-based MAS. causing confusion during the development and execution of LLM-based MAS. For example, a practitioner reported that ``\textit{the documentation lacked examples and guidance for the new multi-agent setup using AgentTool}'' (Adk-Python \#1608).
\end{itemize}
    
\textbf{\emph{Interpretation}}: The \emph{Documentation Issue} category highlights the challenges in LLM-based MAS related to the clarity and completeness of system documentation. These issues include \emph{Unclear Documentation} (14) and \emph{Missing Documentation} (12). Overall, these findings indicate that documentation issues reflect a gap between the system and its supporting documentation in LLM-based MAS.

\subsubsection {Communication Issue}(2.6\%)\label{subsec:CommunicationIssue}
 \emph{Communication Issue} refers to issues related to interaction, message exchange, and information flow between agents in LLM-based MAS that practitioners encounter while developing and using these systems. This category encompasses the following four types.

\begin{itemize}
    \item \textbf{\textsc{Message Format Error}} refers to issues where messages exchanged between agents are incorrectly structured or violate the expected communication protocol (e.g., JSON schema, role fields, or message format), causing failures in inter-agent communication or message processing. For example, a practitioner reported that an entry-point agent sends malformed messages to other agents, requiring a structured format: ``\textit{Entry point agent occasionally sends incorrect or malformed information when communicating with other agents}'' (Agency-Swarm \#175).
    
    \item \textbf{\textsc{Connection Issue}} refers to issues where communication between agents fails when network or connectivity problems occur, such as network errors, incorrect endpoints, or infrastructure constraints, preventing information exchange and coordinated task execution. For example, a practitioner reported that inter-agent communication fails due to a network error: ``\textit{A2A inter-agent communication failure due to network/proxy connectivity problem}'' (Adk-python \#2384).
    
    \item \textbf{\textsc{Role Assignment Issue}} refers to issues where agent roles are incorrectly assigned during inter-agent communication, leading to misinterpretation of context by downstream agents and incorrect outputs. For example, a practitioner reported that: ``\textit{incorrect role assignment in multi-agent mode causes context misinterpretation}'' (Adk-python \#1748).    
    
    \item \textbf{\textsc{Message Transmission Failure}} refers to issues where messages or events are not properly transmitted between agents or system components, leading to incomplete information flow. In such cases, communication signals (e.g., intermediate messages) fail to propagate across system. For example, a practitioner reported that a sub-agent successfully executes a tool, but the corresponding event is not transmitted to the frontend: ``\textit{Sub-agent tool execution event fails to propagate to the frontend}'' (Agno \#3435).
\end{itemize}

\textbf{\emph{Interpretation}}: The \emph{Communication Issue} category highlights key challenges in LLM-based MAS related to message exchange, connectivity, and interaction between agents. The most common issues include \emph{Message Format Error} (12) and \emph{Connection Issue} (6). Overall, these findings indicate that incorrect message structuring, communication failures, and network constraints can disrupt information exchange and interaction between agents in LLM-based MAS.

\subsubsection {Performance Issue} (2.0\%) \label{subsec:PerformanceIssue}
\emph{Performance Issue} refers to issues related to execution efficiency, latency, and resource utilization in LLM-based MAS. This category encompasses the following two types.

\begin{itemize}
    \item \textbf{\textsc{Inefficient Execution}} refers to issues where LLM-based MAS use resources inefficiently during task execution, leading to redundant processing and increased computational or time overhead, such as when unnecessary agents or processes are involved in task execution. For example, a practitioner reported that: ``\textit{GroupChat keeps calling all the available agents and doesn't stop even after output from one agent is generated}'' (AutoGen \#1464).
    
    \item \textbf{\textsc{Response Delay}} refers to issues where LLM-based MAS experience high latency during execution, tool invocation, or response generation. Delays between key stages (e.g., model callbacks, tool calls, or agent communication) result in slow system responsiveness and reduced performance. For example, a practitioner reported that ``\textit{high latency occurs between agent callbacks and tool invocation during multi-agent execution}'' (Adk-python \#1566). 
\end{itemize}

\textbf{\emph{Interpretation}}: The \emph{Performance Issue} category highlights key challenges in LLM-based MAS related to execution efficiency and response latency. The issues include \emph{Inefficient Execution} (9) and \emph{Response Delay} (9). Overall, these findings indicate that redundant processing and execution delays can reduce the efficiency and responsiveness of LLM-based MAS.

\subsubsection {Model Issue} (1.6\%) \label{subsec:ModelIssue}
\emph{Model Issue} refers to issues arising from model behavior and model-level limitations in LLM-based MAS. This category encompasses the following two types.

\begin{itemize}
    \item \textbf{\textsc{Message Format Mismatch}} refers to issues where agents generate outputs in a format that is incompatible with the system’s expected structured output interface. For example, a practitioner reported that the model returns responses in an incorrect object format instead of a properly structured string: ``\textit{Sequential agents with anthropic return the response as an array of strings instead of the expected string format}'' (Flowise \#3792).
    
    \item \textbf{\textsc{API Compatibility Issue}} refers to issues where LLM-based MAS are incompatible with specific model API requirements or provider constraints during model API integration. Model providers enforce strict API rules that may conflict with MAS design, leading to execution failures despite correct system implementation. For example, a practitioner reported that Anthropic’s API enforces strict role-alternation rules that are incompatible with the system’s conversation structure: ``\textit{Anthropic Claude rejects sequential multi-agent messages because its API requires alternating user and assistant roles}'' (Flowise \#3147).
\end{itemize}

\textbf{\emph{Interpretation}}: The \emph{Model Issue} category highlights challenges in LLM-based MAS related to model response formatting and platform-specific constraints. The issues include \emph{Message Format Mismatch} (12) and \emph{API Compatibility Issue} (3). Overall, these findings indicate that model issues arise from mismatches between model outputs, API requirements, and MAS communication requirements.

\subsubsection {Compatibility Issue} (1.4\%) \label{subsec:CompatibilityIssue}
\emph{Compatibility Issue} refers to issues related to version mismatches and incompatibilities between different components or frameworks in LLM-based MAS. This category encompasses the following two types.

\begin{itemize}
    \item \textbf{\textsc{Interface Type Mismatch}} refers to issues where LLM-based MAS produce outputs that do not follow the expected data structures defined by external systems or the expected output formats of model APIs. This leads to violations of required constraints in message content types, field formats, or nested structures. For example, a practitioner reported that ``\textit{sub-agent execution produced array-based content instead of the string format required by the Ollama API}'' (Adk-python \#376).

    \item \textbf{\textsc{API Compatibility Issue}} refers to issues where LLM-based MAS fail to comply with provider-specific API requirements. The request sent to the API does not meet required constraints, leading to runtime rejection despite correct system design. For example, a practitioner reported that ``\textit{crewAI sends conversation messages to WatsonX (IBM’s AI platform) that do not conform to the API's required user/assistant role sequence}'' (CrewAI \#1619).
\end{itemize}

\textbf{\emph{Interpretation}}: The \emph{Compatibility Issue} category highlights challenges in LLM-based MAS related to interface and protocol-level constraints during integration with external services. The issues include \emph{Interface Type Mismatch} (10) and \emph{API Compatibility Issue} (3).  Overall, these findings indicate that compatibility issues arise from violations of external interface schemas and provider-specific conversation rules, leading to integration breakdowns despite correct system design.

\subsubsection {Error Handling Issue} (1.3\%) \label{subsec:ErrorHandlingIssue}
\emph{Error Handling Issue} refers to issues related to failure detection, exception handling, and recovery mechanisms in LLM-based MAS. This category encompasses the following two types.

\begin{itemize}
    \item \textbf{\textsc{Unhandled Exception}} refers to failures where LLM-based MAS do not properly handle runtime exceptions during execution. These issues arise when variables, inputs, or intermediate states are not correctly defined or passed, which leads to unexpected execution failures. For example, a practitioner reported that ``\textit{MetaGPT’s ActionNode.fill function raises an UnboundLocalError when the local variable context is referenced before assignment}'' (MetaGPT \#1839).
    
    \item \textbf{\textsc{Recovery Failure}}  refers to failures where LLM-based MAS are unable to properly restore or correct system state after an error occurs. This happens because partial execution results are not rolled back or cleared. This results in inconsistent system states despite the presence of fallback or recovery mechanisms. For example, a practitioner reported that ``\textit{a guardrail exception fails to remove the prior agent response from the session history}'' (openai-agents-python \#1840).
\end{itemize}

\textbf{\emph{Interpretation}}: The \emph{Error Handling Issue} category highlights challenges in LLM-based MAS related to runtime exception handling and system recovery mechanisms. The issues include \emph{Unhandled Exception} (5) and \emph{Recovery Failure} (7). Overall, these findings indicate that \textit{Error Handling Issues} arise from the inability of systems to properly manage runtime errors and recover from execution breakdowns, leading to reduced reliability and interrupted multi-agent workflows.

\subsubsection {Task Management Issue} (1.2\%) \label{subsec:TaskManagementIssue}
\emph{Task Management Issue} refers to issues arising from task assignment, scheduling, and task execution flow in LLM-based MAS. This category encompasses the following two types.

\begin{itemize}
    \item \textbf{\textsc{Task Allocation Failure}} refers to issues where tasks are not handled by the appropriate agents in a multi-agent workflow, leading to incomplete task execution. For example, a practitioner reported that ``\textit{a `RetrieveUserProxyAgent' agent is sometimes not called when expected, preventing the intended retrieval task from being performed.}'' (AutoGen \#1130).
    
    \item \textbf{\textsc{Task Duplication}} refers to issues where the same task is executed multiple times in a single run, leading to repeated agent executions and duplicated outputs. For example, a practitioner reported that agents are re-invoked after completing their first cycle, causing the pipeline to run again: ``\textit{Agents work twice in every run}'' (CrewAI \#416).
\end{itemize}

\textbf{\emph{Interpretation}}: The \textit{Task Management Issue} category highlights challenges in LLM-based MAS related to task allocation and execution control during multi-agent workflows. The issues include \emph{Task Allocation Failure} (7) and \emph{Task Duplication} (4). Overall, these findings indicate that incorrect task distribution and repeated task execution across agents can reduce coordination effectiveness and lead to inconsistencies in multi-agent workflow execution.

\subsubsection {Security Issue} (0.3\%) \label{subsec:SecurityIssue}
\emph{Security Issue} refers to issues related to access control, data protection, and security vulnerabilities in LLM-based MAS. This category encompasses the following two types.

\begin{itemize}
    \item \textbf{\textsc{Validation Issue}} refers to security issues where insufficient verification of input data leads to unsafe processing, resulting in security vulnerabilities or unintended system behavior. For example, a practitioner reported that \textit{pickle.loads} is used without sufficient input validation, allowing unsafe code execution: ``\textit{incoming agent data is unsafely deserialized using pickle.loads, allowing attackers to execute commands remotely}'' (AgentScope \#635).

    \item \textbf{\textsc{Access Control Issue}} refers to security issues where permission control, or identity-based access mechanisms, are not properly implemented, leading to unauthorized access to user-specific resources. For example, a practitioner reported that ``\textit{wrong user IDs cause agents to return data belonging to other users, resulting in unauthorized access to user-specific resources}'' (Agno \#2279).
\end{itemize}

\textbf{\emph{Interpretation}}: The \emph{Security Issue} category highlights challenges in LLM-based MAS related to input validation and access control in LLM-based MAS. The issues include \emph{Validation Issue} (2) and \emph{Access Control Issue} (1). Overall, these findings indicate that security-related issues mainly arise from insufficient input validation and weak enforcement of access restrictions, which may lead to unsafe processing and unauthorized access in MAS.

\subsection {Category of Causes (RQ2)} \label{subsec:CategoryofCauses}
To address RQ2, we investigate the underlying causes of the issues identified in the final dataset. We categorize the extracted causes into distinct categories to characterize the factors contributing to issues in open-source LLM-based MAS. As mentioned in Section \ref{subsubsec:FormalDataExtraction}, not all closed issues have an associated cause that can be identified. However, an issue may have one or more causes, and all causes identified by the practitioner or development team are recorded. As a result, we identified a total of 540 causes, which were extracted from 57.2\% of the 944 issues and categorized into 14 categories as presented in Table~\ref{tab:category_distribution}. 

\begin{table*}[htbp]
\scriptsize
\caption{Causes of Issues in Open-Source LLM-based Multi-Agent Projects}
\label{tab:category_distribution}
\begin{tabular}{m{4.5cm}m{0.8cm}<{\centering}m{0.8cm}<{\centering}m{9cm}}
\hline
\textbf{Category} & \textbf{Count} & \textbf{\%} & \textbf{Example (Extracted Cause)} \\
\hline

\textit{Workflow Problem (WP)} & 159 & 29.4\% & \textit{System remembers only the first user message and keeps reusing it, leading to wrong answers} (Agno \#3422) \\
\hline

\textit{Tool Integration Problem (TIP)} & 63 & 11.7\% & \textit{Tool-calling agents cannot start managed agents because managed agents are not registered as tools.} (Smolagents \#1353) \\
\hline

\textit{Memory Problem (MP)} & 60 & 11.1\% & \textit{Session history fails to load due to process-bound in-memory caching when teams nest other teams.} (Agno \#4831) \\
\hline

\textit{Integration Problem (IP)} & 56 & 10.4\% & \textit{Integration fails when context does not pass from agent memory to tools, so tools use hardcoded defaults.} (PraisonAI \#289) \\
\hline

\textit{Agent Lifecycle Problem (ALP)} & 56 & 10.4\% & \textit{Initialization fails when an agent name contains a space, as names must be valid Python identifiers.} (smolagents \#640) \\
\hline

\textit{Communication Problem (CP)} & 37 & 6.9\% & \textit{Agents misinterpret each other's messages when responses are incorrectly labeled as user input.} (ADK-Python \#1748) \\
\hline

\textit{Configuration Problem (ConP)} & 34 & 6.3\% & \textit{Configuration fails because the system uses a hardcoded path and ignores user settings.} (Claude-flow \#52) \\
\hline

\textit{Dependency Problem (DP)} & 25 & 4.6\% & \textit{Missing setuptools dependency causes CrewAI import error, breaking multi-agent workflows.} (PraisonAI \#234) \\
\hline

\textit{Documentation Problem (DocP)} & 15 & 2.8\% & \textit{Documentation does not match actual hook implementation, missing events and control flow features.} (Claude-flow \#377) \\
\hline

\textit{Performance Problem (PP)} & 13 & 2.4\% & \textit{Team conversation performance degrades because team sessions keep re-sending old messages.} (Agno \#3793) \\
\hline

\textit{Database Problem (DataP)} & 11 & 2.0\% & \textit{Missing tables in database schema break state persistence.} (Claude-flow \#192) \\
\hline

\textit{Infrastructure Problem (InfP)} & 6 & 1.1\% & \textit{Kubernetes networking constraints prevent pods from reaching their own external URL.} (Activepieces \#9038) \\
\hline

\textit{Frontend Problem (FP)} & 3 & 0.6\% & \textit{Agent outputs are truncated when fixed-position boxes run out of space} (Agno \#3680) \\
\hline

\textit{Security Problem (SP)} & 2 & 0.4\% & \textit{The function loads incoming untrusted data in a way that can run hidden malicious code.} (AgentScope \#635) \\
\hline
\end{tabular}
\end{table*}

\subsubsection {Results}
The results show that the most prominent cause is \textit{Workflow Problem (WP)} (29.4\%), followed by \textit{Tool Integration Problem (TIP)} (11.7\%), and \textit{Memory Problem (MP)} (11.1\%). Examples of each category of cause, along with their counts and proportions, are also provided in Table~\ref{tab:category_distribution}. It should be noted that certain categories of issues can potentially serve as causes of other issues. For example, problems in the \textit{Agent Lifecycle Problem (ALP)} category can disrupt workflow execution, leading to problems in the \textit{Workflow Problem (WP)} category. Such failures may also disrupt inter-agent coordination and message exchange during execution, resulting in problems in the \textit{Communication Problem (CP)} category.

\begin{itemize}
    \item \textbf{\textsc{Workflow Problem (WP)}} refers to problems in workflow execution and task coordination in LLM-based MAS. These problems are caused by limitations in workflow design and execution, leading to incorrect, incomplete, or failed outcomes. It encompasses eleven types: (1) \textsc{Context Management Problem}: Problems arising from failures in transferring task-related context (e.g., instructions, communication history, or execution state) between agents during workflow execution, causing incorrect coordination or task outcomes. (2) \textsc{Coordination Problem}: Problems arising from weak interaction and synchronization among multiple agents. (3) \textsc{Error Transmission Problem}: Problems arising when failures in an earlier workflow step affect subsequent execution stages, causing incorrect task outcomes. 
    
    (4) \textsc{Handoff Data Problem}: Problems arising when task-related data, execution context, or required parameters are missing or incompletely transferred during a handoff between agents. For example, a user reported that ``\textit{required execution events were incorrectly classified and filtered out during the handoff}'' (Adk-Python \#2073). (5) \textsc{Loop Problem}: Problems arising from missing or incorrect termination conditions or flawed control logic. (6) \textsc{Message Routing Problem}: Problems arising due to failures in processing, forwarding, or delivering messages between agents during execution. 
    
    (7) \textsc{Orchestration Problem}: Problems arising from improper management of task dependencies and workflow execution across multiple agents, affecting overall execution flow. (8) \textsc{Resource Contention Problem}: Problems arising when multiple agents compete for limited shared resources (e.g., memory, files, tools, or execution slots) during concurrent execution, causing blocking, conflicts, or delayed task completion. (9) \textsc{Sequential Execution Problem}: Problems arising due to incorrect ordering or dependency conflicts in stepwise execution. (10) \textsc{Workflow State Error}: Problems arising from incorrect or inconsistent system state during a single workflow run, causing agents to act on outdated or corrupted intermediate data across sequential steps. (11) \textsc{External Tool Problem}: Problems arising when an external tool itself fails or produces unexpected outputs during workflow execution, despite being correctly integrated and invoked, resulting in incorrect or incomplete workflow outcomes.
    
    \item \textbf{\textsc{Tool Integration Problem (TIP)}} refers to problems concerning the integration and interaction of external tools in LLM-based MAS. These problems are caused by failures in integrating, invoking, or communicating with tools, including parameter handling, tool access, or schema consistency, leading to invocation failures, incorrect data transmission, or incomplete task outcomes. It encompasses four types: (1) \textsc{MCP Integration Problem}: Problems arising due to failures in integrating or communicating with external Model Context Protocol (MCP)-based systems. (2) \textsc{Parameter Integration Problem}: Problems arising due to required input parameters not being correctly passed to tools during invocation. (3) \textsc{Tool Execution Problem}: Problems arising due to failures in accessing or executing tools, where the tool cannot run as intended (e.g., resource locks, runtime crashes, timeouts). For example, a user explained that ``\textit{Claude Code tools were already active or locked in the current session, preventing Claude-Flow from accessing them}'' (Claude-Flow \#59). (4) \textsc{Tool Schema Problem}: Problems arising due to mismatches or inconsistencies between tool definitions and expected input or output schemas.
    
    \item \textbf{\textsc{Memory Problem (MP)}} refers to problems concerning the storage, retrieval, and utilization of memory in LLM-based MAS. These problems arise due to limitations in memory mechanisms or contextual consistency, leading to consistency errors and incorrect responses across sessions. It encompasses three types: (1) \textsc{Memory Persistence Problem}: Problems arising from failures in storing, retrieving, or retaining contextual information across execution steps, causing agents to lose previously established knowledge or states. (2) \textsc{Memory Management Problem}: Problems arising due to failures in storing or updating internal or external memory, affecting memory consistency and correctness. For example, a user explained that ``\textit{external memory storage failed because agent roles were appended to the collection name, exceeding the allowed length}'' (CrewAI \#2985). (3) \textsc{State Management Problem}: Problems arising due to inconsistent or incorrect maintenance of system state across interactions or sessions.
    
    \item \textbf{\textsc{Integration Problem (IP)}} refers to problems concerning the integration and compatibility of components in LLM-based MAS. These problems arise due to failures in component compatibility, coordination among agents, or configuration alignment across agents, models, and execution frameworks, leading to communication inconsistencies, configuration mismatches, and incorrect system behavior across workflows. It encompasses two types: (1) \textsc{Agent Integration Problem}: Problems arising due to failures in integrating agents within hierarchical MAS. For example, a user reported that ``\textit{the issue occurred because the speaker-selection agent was not properly integrated with the custom model client}'' (AutoGen \#2956). (2) \textsc{Parameter Propagation Problem}: Problems arising when required parameters or execution settings are not correctly transferred between interacting components in LLM-based MAS.
    
    \item \textbf{\textsc{Agent Lifecycle Problem (ALP)}} refers to problems concerning the lifecycle of agents in LLM-based MAS, including their creation, initialization, and execution. These problems arise from failures in agent initialization or runtime execution, leading to improper agent setup, execution failures, or inconsistent agent behavior. It encompasses two types: (1) \textsc{Agent Creation Problem}: Problems arising due to created agents not being properly initialized with required configurations or model clients. For example, a user reported that ``\textit{a custom model client failed in a group chat because the internal speaker-selection agent was initialized without registering the custom model client}'' (AutoGen \#2643). (2) \textsc{Agent Runtime Problem}: Problems arising when an agent fails to execute an assigned task correctly due to runtime errors or execution interruptions. 

    \item \textbf{\textsc{Communication Problem (CP)}} refers to problems concerning the exchange of information between agents in LLM-based MAS. These problems arise due to failures in message passing, routing, or maintaining shared context among agents, leading to incorrect information transfer and disrupted execution continuity across the workflow. It encompasses three types: (1) \textsc{Handoff Problem}: Problems arising during the transfer of control, responsibility, or task ownership between agents, where the receiving agent fails to correctly continue execution. For example, a user reported that ``\textit{a handoff between agents returned no task because the user message was added to the chat history after the LLM processed the handoff tool, making the message unavailable to the receiving agent}'' (PraisonAI \#795). (2) \textsc{Inter-Agent Communication Problem}: Problems arising due to failures in delivering or routing messages between agents in LLM-based MAS. (3) \textsc{Message Handling Problem}: Problems arising due to incorrect processing, interpretation, or filtering of received messages within agents.
    
    \item \textbf{\textsc{Configuration Problem (ConP)}} refers to problems related to the setup and specification of parameters and system configurations in LLM-based MAS. These problems arise due to missing or incorrect configuration of models, agents, or execution frameworks, leading to initialization failures and runtime errors. It encompasses the following two types: (1) \textsc{Setup Problem}: Problems arising due to missing, incompatible, or incorrectly initialized configuration files or environment dependencies during MAS or agent setup. (2) \textsc{Parameter Problem}: Problems arising due to required parameters, model settings, or execution arguments being incorrectly specified or propagated. For example, a user reported that ``\textit{CaptainAgent failed because the internal AgentBuilder ignored the explicitly provided LLMConfig parameter and instead looked for a legacy OAI\_CONFIG\_LIST file, causing a file-not-found error}'' (Ag2 \#1446).
    
    \item \textbf{\textsc{Dependency Problem (DP)}} refers to problems concerning incompatibilities among system dependencies in LLM-based MAS. These problems arise due to mismatched, missing, or incompatible external APIs or modules, leading to execution failures or runtime errors. It encompasses three types: (1) \textsc{API Compatibility Problem}: Problems arising due to incompatible or changed API interfaces. (2) \textsc{Import Problem}: Problems arising due to missing or improperly loaded modules, files, or configuration imports. (3) \textsc{Version Compatibility Problem}: Problems arising due to mismatched or incompatible software versions following upgrades or changes in the execution environment. For example, a user reported that ``\textit{upgrading Flowise to v3.0.3 caused existing AgentFlow v1 flows to fail due to incompatibility between the updated dependencies and the existing flow configuration}'' (Flowise \#4756).
     
    \item \textbf{\textsc{Documentation Problem (DocP)}} refers to problems concerning incomplete or inconsistent documentation in LLM-based MAS. These problems arise due to missing, outdated, or inconsistent documentation, leading to usability issues, configuration errors, or incorrect system execution. It encompasses two types: (1) \textsc{Missing Documentation}: Problems arising due to required information not being provided in documentation. (2) \textsc{Outdated Documentation}: Problems arising due to documentation not being updated after system changes. For example, a user reported that a ``\textit{sub-agent execution failed because the documentation incorrectly instructed hooks notification while the actual command expected hooks notify}'' (Claude-flow \#504).
    
    \item \textbf{\textsc{Performance Problem (PP)}} refers to problems concerning the efficiency and speed of LLM-based MAS. These problems arise due to limitations in parallel execution, scalability, timeouts, or processing latency, leading to slow execution or high resource usage. It encompasses four types: (1) \textsc{Parallel Execution Problem}: Problems arising due to failures in concurrent execution across agents. (2) \textsc{Scalability Problem}: Problems arising due to reduced performance as system size or workload increases.(3) \textsc{Timeout Problem}: Problems arising when operations exceed the allowed execution time. (4) \textsc{Latency Problem}: Problems arising due to delays caused by inefficient processing or large data operations. For example, a user explained that ``\textit{shared memory access became slow as the dataset grew because load() used deepcopy() on every call, and duplicate records led to increased memory usage, resulting in high latency}'' (Camel \#3185).

    \item \textbf{\textsc{Database Problem (DataP)}} refers to problems concerning database structure, initialization, data access, and schema consistency in LLM-based MAS. These problems arise due to failures in database setup, data retrieval, or schema alignment, leading to execution failures or incorrect system behavior. It encompasses three types: (1) \textsc{Database Initialization Failure}: Problems arising due to errors during database creation or setup. (2) \textsc{Data Retrieval Failure}: Problems arising due to failure in accessing or retrieving required data. (3) \textsc{Database Schema Inconsistency}: Problems arising due to mismatches between expected and actual database schema. For example, a user reported that ``\textit{agent creation failed because the agents table lacked the role column that the application expected during data insertion}'' (Claude-Flow \#403).
    
    \item \textbf{\textsc{Infrastructure Problem (InfP)}} refers to problems concerning the technical infrastructure, execution environment, and deployment of LLM-based MAS. These problems arise due to failures or limitations in system infrastructure, deployment environments, or file handling, leading to execution failures or system unavailability. It encompasses three types: (1) \textsc{Server Problem}: Problems arising from server configuration, availability, or resource limitations. (2) \textsc{File System Handling Problem}: Problems arising from file access, storage, or path-related issues. (3) \textsc{Deployment Problem}: Problems arising from infrastructure or runtime environment issues during system deployment. For example, a user explained that ``\textit{AI agents failed to be created in a Kubernetes deployment because network policies prevented pods from reaching external URLs, causing connection timeouts to external AI providers}'' (Activepieces \#9038).

    \item \textbf{\textsc{Frontend Problem (FP)}} refers to problems concerning the user interface and output handling of LLM-based MAS. These problems arise due to failures or limitations in user interface rendering, event propagation, or output display, leading to incorrect, incomplete, or unavailable information being presented to users. It encompasses two types: (1) \textsc{User Interface Problem}: Problems arising from failures in rendering or presenting the user interface during MAS execution. (2) \textsc{Output Problem}: Problems arising from failures or limitations in displaying, updating, or streaming agent outputs. For example, a user explained that ``\textit{streaming agent outputs are truncated when fixed-position boxes run out of available display space, preventing the display from scrolling to the latest output}'' (Agno \#3680).
    
    \item \textbf{\textsc{Security Problem (SP)}} refers to problems concerning vulnerabilities that expose LLM-based MAS to unauthorized access, malicious execution, or unauthorized manipulation of system control. These problems arise due to weaknesses in input validation, deserialization, or function access control, leading to system compromise or unauthorized actions. It encompasses two types: (1) \textsc{Prompt Injection Problem}: Problems arising due to manipulation of model inputs to alter system behavior or bypass constraints. (2) \textsc{Insecure Deserialization Problem}: Problems arising due to unsafe deserialization of malicious data, potentially allowing unauthorized code execution. For example, a user explained that ``\textit{the system unsafely deserialized incoming agent data using a vulnerable deserialization method, allowing malicious code execution}'' (Agentscope \#635).   
\end{itemize}

\subsubsection{Causes to Issues Mapping}\label{sec:CauseMapping} 
Table~\ref{tab:issues-causes-mapping} illustrates the mapping between \textit{Issue} categories and \textit{Cause} categories identified in the development and use of open-source LLM-based MAS, along with their distributions. To improve readability, abbreviations are used to denote each \textit{Cause} category for instance, “WP” refers to \textit{Workflow Problem}. The full names of all \textit{Cause} categories are provided in the note of Table~\ref{tab:issues-causes-mapping}.

Over three quarters, 87.0\% of \textit{Implementation Issue} are associated with identified causes, representing the highest rate among all issue categories. These issues are mainly caused by \textit{Workflow Problem (WP)} (24).

Only 13.6\% of \textit{Feature Request} cases are associated with identified causes, representing the lowest rate among all issue categories. \textit{Workflow Problem (WP)} (7) is identified as the most frequent cause.

The largest category among all issues is \textit{Orchestration \& Execution Issue}. 54.8\% of these issues are associated with identified causes, and the most frequent causes are \textit{Workflow Problem (WP)} (65), \textit{Agent Lifecycle Problem (ALP)} (10), and \textit{Integration Problem (IP)} (9).

For \textit{Error Handling Issue}, causes are identified in 83.3\% of cases. These issues mainly originate from \textit{Tool Integration Problem (TIP)} (3).

For \textit{Tool Issue}, 73.6\% of cases are associated with identified causes. These issues mainly originate from \textit{Tool Integration Problem (TIP)} (23) and \textit{Workflow Problem (WP)} (18).

For \textit{Integration Issue}, causes are identified in 71.4\% of cases, mainly arising from \textit{Integration Problem (IP)} (17).

For \textit{Environment \& Configuration Issue}, causes are identified in 69.7\% of cases, mainly arising from \textit{Dependency Problem (DP)} (7) and \textit{Configuration Problem (ConP)} (7).

For \textit{Compatibility Issue}, causes are identified in 69.2\% of cases, mainly arising from \textit{Communication Problem (CP)} (3).

For \textit{Model Issue}, causes are identified in 66.7\% of cases, mainly originating from \textit{Workflow Problem (WP)} (4).

For \textit{Security Issue}, causes are identified in 66.7\% of cases, mainly originating from \textit{Security Problem (SP)} (2).

For \textit{Memory \& State Management Issue}, causes are identified in 60.4\% of cases. These issues mainly originate from \textit{Memory Problem (MP)} (43).

For \textit{Communication Issue}, causes are identified in 64.0\% of cases. These issues mainly originate from \textit{Communication Problem (CP)} (11).

For \textit{Role Issue}, causes are identified in 52.8\% of cases. These issues are mainly caused by \textit{Workflow Problem (WP)} (7).

For \textit{Documentation Issue}, causes are identified in 57.7\% of cases. These issues mainly originate from \textit{Documentation Problem (DocP)} (10).

For \textit{Output Issue}, causes are identified in 48.3\% of cases. These issues are mainly caused by \textit{Integration Problem (IP)} (3), \textit{Workflow Problem (WP)} (3), and \textit{Frontend Problem (FP)} (3).

For \textit{Performance Issue}, causes are identified in 36.8\% of cases. These issues are mainly caused by \textit{Memory Problem (MP)} (2), \textit{Workflow Problem (WP)} (2), and \textit{Performance Problem (PP)} (2).

For \textit{Task Management Issue}, causes are identified in 36.4\% of cases. These issues are mainly caused by \textit{Workflow Problem (WP)} (4).

\begin{table*}[h]
\scriptsize
\caption{Mapping between issue categories (vertical) and cause categories (horizontal)}
\label{tab:issues-causes-mapping}
\begin{adjustbox}{center}
\begin{tabular}{ccccccccccccccc}
                                                    & \multicolumn{1}{c}{\textbf{WP}} & \multicolumn{1}{c}{\textbf{TIP}} & \multicolumn{1}{c}{\textbf{MP}} 
                                                    & \multicolumn{1}{c}{\textbf{IP}} & \multicolumn{1}{c}{\textbf{ALP}} & \multicolumn{1}{c}{\textbf{CP}} 
                                                    & \multicolumn{1}{c}{\textbf{ConP}} & \multicolumn{1}{c}{\textbf{DP}} & \multicolumn{1}{c}{\textbf{DocP}} 
                                                    & \multicolumn{1}{c}{\textbf{PP}} & \multicolumn{1}{c}{\textbf{DataP}} & \multicolumn{1}{c}{\textbf{InfP}} 
                                                    & \multicolumn{1}{c}{\textbf{FP}} & \multicolumn{1}{c}{\textbf{SP}} \\
\rowcolor[HTML]{acddde} \textbf{Orchestration \& Execution Issue} 
                                                    & \cellcolor[HTML]{5da6d6}65      & \cellcolor[HTML]{aad4ed}7       & \cellcolor[HTML]{e9f7ff}3 
                                                    & \cellcolor[HTML]{aad4ed}9       & \cellcolor[HTML]{cce8f7}10      & \cellcolor[HTML]{aad4ed}8 
                                                    & \cellcolor[HTML]{cce8f7}4       & \cellcolor[HTML]{e9f7ff}1       & \cellcolor[HTML]{f4f8fb}0
                                                    & \cellcolor[HTML]{cce8f7}5       & \cellcolor[HTML]{e9f7ff}1       & \cellcolor[HTML]{e9f7ff}1
                                                    & \cellcolor[HTML]{f4f8fb}0       & \cellcolor[HTML]{f4f8fb}0 \\
\rowcolor[HTML]{e2f8dc} \textbf{Memory \& State Management Issue} 
                                                    & \cellcolor[HTML]{82bde2}16      & \cellcolor[HTML]{e9f7ff}1       & \cellcolor[HTML]{5da6d6}43 
                                                    & \cellcolor[HTML]{e9f7ff}1       & \cellcolor[HTML]{aad4ed}6       & \cellcolor[HTML]{cce8f7}2 
                                                    & \cellcolor[HTML]{e9f7ff}3       & \cellcolor[HTML]{e9f7ff}0       & \cellcolor[HTML]{f4f8fb}0
                                                    & \cellcolor[HTML]{e9f7ff}3       & \cellcolor[HTML]{aad4ed}8      & \cellcolor[HTML]{e9f7ff}1       
                                                    & \cellcolor[HTML]{f4f8fb}0       & \cellcolor[HTML]{f4f8fb}0 \\
\rowcolor[HTML]{acddde} \textbf{Feature Request} 
                                                    & \cellcolor[HTML]{aad4ed}7       & \cellcolor[HTML]{e9f7ff}2        & \cellcolor[HTML]{f4f8fb}0 
                                                    & \cellcolor[HTML]{e9f7ff}2       & \cellcolor[HTML]{e9f7ff}1        & \cellcolor[HTML]{e9f7ff}2 
                                                    & \cellcolor[HTML]{e9f7ff}1       & \cellcolor[HTML]{e9f7ff}0        & \cellcolor[HTML]{e9f7ff}1       
                                                    & \cellcolor[HTML]{e9f7ff}1       & \cellcolor[HTML]{f4f8fb}0        & \cellcolor[HTML]{f4f8fb}0
                                                    & \cellcolor[HTML]{f4f8fb}0       & \cellcolor[HTML]{f4f8fb}0 \\
\rowcolor[HTML]{e2f8dc} \textbf{Implementation Issue} 
                                                    & \cellcolor[HTML]{5da6d6}24      & \cellcolor[HTML]{aad4ed}7        & \cellcolor[HTML]{cce8f7}4 
                                                    & \cellcolor[HTML]{aad4ed}11      & \cellcolor[HTML]{aad4ed}18       & \cellcolor[HTML]{e9f7ff}1 
                                                    & \cellcolor[HTML]{aad4ed}12      & \cellcolor[HTML]{cce8f7}5        & \cellcolor[HTML]{cce8f7}2
                                                    & \cellcolor[HTML]{e9f7ff}0       & \cellcolor[HTML]{e9f7ff}2        & \cellcolor[HTML]{e9f7ff}1
                                                    & \cellcolor[HTML]{f4f8fb}0       & \cellcolor[HTML]{f4f8fb}0 \\
\rowcolor[HTML]{acddde} \textbf{Tool Issue} 
                                                    & \cellcolor[HTML]{82bde2}18      & \cellcolor[HTML]{5da6d6}23       & \cellcolor[HTML]{e9f7ff}3 
                                                    & \cellcolor[HTML]{cce8f7}4       & \cellcolor[HTML]{e9f7ff}4        & \cellcolor[HTML]{cce8f7}4 
                                                    & \cellcolor[HTML]{e9f7ff}1       & \cellcolor[HTML]{e9f7ff}4        & \cellcolor[HTML]{cce8f7}1
                                                    & \cellcolor[HTML]{e9f7ff}1       & \cellcolor[HTML]{e9f7ff}0        & \cellcolor[HTML]{e9f7ff}1
                                                    & \cellcolor[HTML]{f4f8fb}0       & \cellcolor[HTML]{f4f8fb}0 \\
\rowcolor[HTML]{e2f8dc} \textbf{Integration Issue} 
                                                    & \cellcolor[HTML]{e9f7ff}2       & \cellcolor[HTML]{82bde2}11       & \cellcolor[HTML]{e9f7ff}1 
                                                    & \cellcolor[HTML]{5da6d6}17      & \cellcolor[HTML]{e9f7ff}3        & \cellcolor[HTML]{e9f7ff}2 
                                                    & \cellcolor[HTML]{e9f7ff}3       & \cellcolor[HTML]{cce8f7}5        & \cellcolor[HTML]{e9f7ff}1
                                                    & \cellcolor[HTML]{f4f8fb}0       & \cellcolor[HTML]{f4f8fb}0       & \cellcolor[HTML]{f4f8fb}0       
                                                    & \cellcolor[HTML]{f4f8fb}0       & \cellcolor[HTML]{f4f8fb}0 \\
\rowcolor[HTML]{acddde} \textbf{Role Issue} 
                                                    & \cellcolor[HTML]{aad4ed}7       & \cellcolor[HTML]{f4f8fb}0        & \cellcolor[HTML]{f4f8fb}0 
                                                    & \cellcolor[HTML]{cce8f7}5       & \cellcolor[HTML]{aad4ed}6        & \cellcolor[HTML]{f4f8fb}0 
                                                    & \cellcolor[HTML]{e9f7ff}1       & \cellcolor[HTML]{f4f8fb}0        & \cellcolor[HTML]{f4f8fb}0
                                                    & \cellcolor[HTML]{f4f8fb}0       & \cellcolor[HTML]{f4f8fb}0       & \cellcolor[HTML]{f4f8fb}0
                                                    & \cellcolor[HTML]{f4f8fb}0       & \cellcolor[HTML]{f4f8fb}0 \\
\rowcolor[HTML]{e2f8dc} \textbf{Environment \& Configuration Issue} 
                                                    & \cellcolor[HTML]{e9f7ff}1       & \cellcolor[HTML]{e9f7ff}2        & \cellcolor[HTML]{f4f8fb}0 
                                                    & \cellcolor[HTML]{e9f7ff}1       & \cellcolor[HTML]{e9f7ff}3        & \cellcolor[HTML]{f4f8fb}0 
                                                    & \cellcolor[HTML]{aad4ed}7       & \cellcolor[HTML]{aad4ed}7        & \cellcolor[HTML]{f4f8fb}0       
                                                    & \cellcolor[HTML]{f4f8fb}0       & \cellcolor[HTML]{f4f8fb}0        & \cellcolor[HTML]{e9f7ff}2       
                                                    & \cellcolor[HTML]{f4f8fb}0       & \cellcolor[HTML]{f4f8fb}0 \\
\rowcolor[HTML]{acddde} \textbf{Output Issue} 
                                                    & \cellcolor[HTML]{e9f7ff}3       & \cellcolor[HTML]{e9f7ff}1        & \cellcolor[HTML]{e9f7ff}1 
                                                    & \cellcolor[HTML]{e9f7ff}3       & \cellcolor[HTML]{e9f7ff}2        & \cellcolor[HTML]{e9f7ff}1 
                                                    & \cellcolor[HTML]{f4f8fb}0       & \cellcolor[HTML]{f4f8fb}0        & \cellcolor[HTML]{f4f8fb}0       
                                                    & \cellcolor[HTML]{f4f8fb}0       & \cellcolor[HTML]{f4f8fb}0        & \cellcolor[HTML]{f4f8fb}0       
                                                    & \cellcolor[HTML]{e9f7ff}3       & \cellcolor[HTML]{f4f8fb}0 \\
\rowcolor[HTML]{e2f8dc} \textbf{Documentation Issue} 
                                                    & \cellcolor[HTML]{f4f8fb}0       & \cellcolor[HTML]{e9f7ff}2        & \cellcolor[HTML]{e9f7ff}1 
                                                    & \cellcolor[HTML]{f4f8fb}0       & \cellcolor[HTML]{f4f8fb}0        & \cellcolor[HTML]{e9f7ff}0 
                                                    & \cellcolor[HTML]{e9f7ff}1       & \cellcolor[HTML]{f4f8fb}1        & \cellcolor[HTML]{aad4ed}10       
                                                    & \cellcolor[HTML]{f4f8fb}0       & \cellcolor[HTML]{f4f8fb}0        & \cellcolor[HTML]{f4f8fb}0       
                                                    & \cellcolor[HTML]{f4f8fb}0       & \cellcolor[HTML]{f4f8fb}0 \\
\rowcolor[HTML]{acddde} \textbf{Communication Issue} 
                                                    & \cellcolor[HTML]{e9f7ff}3       & \cellcolor[HTML]{e9f7ff}1        & \cellcolor[HTML]{f4f8fb}0 
                                                    & \cellcolor[HTML]{e9f7ff}1       & \cellcolor[HTML]{f4f8fb}0        & \cellcolor[HTML]{aad4ed}11 
                                                    & \cellcolor[HTML]{f4f8fb}0       & \cellcolor[HTML]{f4f8fb}0        & \cellcolor[HTML]{f4f8fb}0       
                                                    & \cellcolor[HTML]{f4f8fb}0       & \cellcolor[HTML]{f4f8fb}0        & \cellcolor[HTML]{f4f8fb}0
                                                    & \cellcolor[HTML]{f4f8fb}0       & \cellcolor[HTML]{f4f8fb}0 \\
\rowcolor[HTML]{e2f8dc} \textbf{Performance Issue} 
                                                    & \cellcolor[HTML]{e9f7ff}2       & \cellcolor[HTML]{e9f7ff}1        & \cellcolor[HTML]{e9f7ff}2 
                                                    & \cellcolor[HTML]{f4f8fb}0       & \cellcolor[HTML]{f4f8fb}0        & \cellcolor[HTML]{f4f8fb}0 
                                                    & \cellcolor[HTML]{f4f8fb}0       & \cellcolor[HTML]{f4f8fb}0        & \cellcolor[HTML]{f4f8fb}0
                                                    & \cellcolor[HTML]{e9f7ff}2       & \cellcolor[HTML]{f4f8fb}0        & \cellcolor[HTML]{f4f8fb}0
                                                    & \cellcolor[HTML]{f4f8fb}0       & \cellcolor[HTML]{f4f8fb}0 \\
\rowcolor[HTML]{acddde} \textbf{Model Issue} 
                                                    & \cellcolor[HTML]{cce8f7}4       & \cellcolor[HTML]{e9f7ff}1        & \cellcolor[HTML]{f4f8fb}0 
                                                    & \cellcolor[HTML]{e9f7ff}2       & \cellcolor[HTML]{f4f8fb}0        & \cellcolor[HTML]{e9f7ff}3 
                                                    & \cellcolor[HTML]{f4f8fb}0       & \cellcolor[HTML]{f4f8fb}0        & \cellcolor[HTML]{f4f8fb}0
                                                    & \cellcolor[HTML]{f4f8fb}0       & \cellcolor[HTML]{f4f8fb}0        & \cellcolor[HTML]{f4f8fb}0
                                                    & \cellcolor[HTML]{f4f8fb}0       & \cellcolor[HTML]{f4f8fb}0 \\
\rowcolor[HTML]{e2f8dc} \textbf{Compatibility Issue} 
                                                    & \cellcolor[HTML]{e9f7ff}1       & \cellcolor[HTML]{e9f7ff}1        & \cellcolor[HTML]{f4f8fb}0 
                                                    & \cellcolor[HTML]{f4f8fb}0       & \cellcolor[HTML]{e9f7ff}1        & \cellcolor[HTML]{e9f7ff}3 
                                                    & \cellcolor[HTML]{e9f7ff}1       & \cellcolor[HTML]{e9f7ff}2        & \cellcolor[HTML]{f4f8fb}0
                                                    & \cellcolor[HTML]{f4f8fb}0       & \cellcolor[HTML]{f4f8fb}0        & \cellcolor[HTML]{f4f8fb}0
                                                    & \cellcolor[HTML]{f4f8fb}0       & \cellcolor[HTML]{f4f8fb}0 \\
\rowcolor[HTML]{acddde} \textbf{Error Handling Issue} 
                                                    & \cellcolor[HTML]{e9f7ff}2       & \cellcolor[HTML]{e9f7ff}3        & \cellcolor[HTML]{e9f7ff}2 
                                                    & \cellcolor[HTML]{f4f8fb}0       & \cellcolor[HTML]{e9f7ff}2        & \cellcolor[HTML]{f4f8fb}0 
                                                    & \cellcolor[HTML]{f4f8fb}0       & \cellcolor[HTML]{f4f8fb}0        & \cellcolor[HTML]{f4f8fb}0
                                                    & \cellcolor[HTML]{e9f7ff}1       & \cellcolor[HTML]{f4f8fb}0        & \cellcolor[HTML]{f4f8fb}0
                                                    & \cellcolor[HTML]{f4f8fb}0       & \cellcolor[HTML]{f4f8fb}0 \\
\rowcolor[HTML]{e2f8dc} \textbf{Task Management Issue} 
                                                    & \cellcolor[HTML]{cce8f7}4       & \cellcolor[HTML]{f4f8fb}0        & \cellcolor[HTML]{f4f8fb}0 
                                                    & \cellcolor[HTML]{f4f8fb}0       & \cellcolor[HTML]{f4f8fb}0        & \cellcolor[HTML]{f4f8fb}0 
                                                    & \cellcolor[HTML]{f4f8fb}0       & \cellcolor[HTML]{f4f8fb}0        & \cellcolor[HTML]{f4f8fb}0
                                                    & \cellcolor[HTML]{f4f8fb}0       & \cellcolor[HTML]{f4f8fb}0        & \cellcolor[HTML]{f4f8fb}0
                                                    & \cellcolor[HTML]{f4f8fb}0       & \cellcolor[HTML]{f4f8fb}0 \\
\rowcolor[HTML]{acddde} \textbf{Security Issue} 
                                                    & \cellcolor[HTML]{f4f8fb}0       & \cellcolor[HTML]{f4f8fb}0        & \cellcolor[HTML]{f4f8fb}0 
                                                    & \cellcolor[HTML]{f4f8fb}0       & \cellcolor[HTML]{f4f8fb}0        & \cellcolor[HTML]{f4f8fb}0 
                                                    & \cellcolor[HTML]{f4f8fb}0       & \cellcolor[HTML]{f4f8fb}0        & \cellcolor[HTML]{f4f8fb}0
                                                    & \cellcolor[HTML]{f4f8fb}0       & \cellcolor[HTML]{f4f8fb}0        & \cellcolor[HTML]{f4f8fb}0
                                                    & \cellcolor[HTML]{f4f8fb}0       & \cellcolor[HTML]{e9f7ff}2 \\
\end{tabular}
\end{adjustbox}
\begin{minipage}{\textwidth}
\vspace{0.1cm}
\vspace{0.1cm}
    \textbf{Full names of every cause category abbreviation}: WP: \textit{Workflow Problem}; TIP: \textit{Tool Integration Problem};  MP: \textit{Memory Problem}; IP: \textit{Integration Problems}; ALP: \textit{Agent Lifecycle Problem}; CP: \textit{Communication Problem};
     ConP: \textit{Configuration Problem}; DP: \textit{Dependency Problem}; DocP: \textit{Documentation Problem}; PP: \textit{Performance Problem};  DataP: \textit{Database Problem}; InfP: \textit{Infrastructure Problem}; FP: \textit{Frontend Problem} ; SP: \textit{Security Problem}.
\end{minipage}
\end{table*}

\subsubsection{Interpretation}\label{sec:interpretation} 
In this section, we analyze the frequency and distribution of different causes across various issue categories in LLM-based MAS. We further map these causes to specific issue types to identify the dominant root causes associated with different issues.

\textbf{\emph{Frequency of the causes}}: The frequency of the causes shows that \textit{Workflow Problem (WP)} is the most frequent cause of issues, as the results indicate. Since LLM-based MAS rely on sequential execution and shared context, most failures arise from incorrect workflow state handling and repeated use of outdated context, leading to incorrect outputs. This highlights workflow consistency as a key challenge in multi-agent coordination. \textit{Tool Integration Problem (TIP)} and \textit{Memory Problem (MP)} also significantly contribute to issues in LLM-based MAS. TIP reflects failures in tool registration or invocation, where agents cannot access required external tools. \textit{Memory Problem (MP)} indicates failures in retaining or retrieving contextual information across agent interactions, causing agents in LLM-based MAS to lose previously established context during task execution. Together, \textit{Tool Integration Problem (TIP)} and \textit{Memory Problem (MP)} represent important causes of issues in LLM-based MAS, particularly those related to tool integration and memory management. Additionally, \textit{Integration Problem (IP)}, \textit{Agent Lifecycle Problem (ALP)}, \textit{Communication Problem (CP)}, and \textit{Configuration Problem (ConP)} appear with moderate frequency. These causes mainly concern system setup and coordination, including incorrect configurations, naming constraints, and message misinterpretation. Although less frequent than WP, these causes reflect important challenges in configuring and coordinating LLM-based MAS. The remaining categories, including \textit{Dependency Problem (DP)}, \textit{Documentation Problem (DocP)}, \textit{Performance Problem (PP)}, \textit{Database Problem (DataP)}, \textit{Infrastructure Problem (InfP)}, \textit{Frontend Problem (FP)} and \textit{Security Problem (SP)}, occur less frequently but still impact system stability, usability, scalability, and security of LLM-based MAS. Overall, most causes are concentrated in workflow execution, tool usage, and memory handling, while the remaining categories represent other underlying causes in LLM-based MAS.

\textbf{\emph{Mapping of causes to issues}}: \textit{Workflow Problem (WP)} is the most frequent cause across multiple issue categories. 57.0\% of \textit{Orchestration \& Execution Issue} causes originate from \textit{Workflow Problem (WP)}, reflecting the central role of workflow design in agent coordination and task execution. Similarly, 51.2\% of \textit{Memory \& State Management Issue} cases are caused by \textit{Memory Problem (MP)}, highlighting the close relationship between workflow execution and memory handling when agents lose context or state updates are not properly propagated. \textit{Tool Integration Problem (TIP)} is the dominant cause of \textit{Tool Issue}, accounting for 35.9\% of cases. These issues typically arise from tool execution failures, invocation errors, and schema mismatches. \textit{Integration Problem (IP)} is the primary cause of \textit{Integration Issue}, accounting for 37.7\% of cases, reflecting the challenges of coordinating multiple components, APIs, and frameworks within the MAS. Overall, \textit{Workflow Problem (WP)} and \textit{Tool Integration Problem (TIP)} are the most dominant root causes of issues in LLM-based MAS. This highlights the importance of robust workflow design and reliable tool integration in LLM-based MAS.

\subsection {Category of Solutions (RQ3)}\label{subsec:CategoryofSolutions}
To address RQ3, we investigate the solutions used to resolve the identified issues in open-source LLM-based MAS. We categorize the identified solutions into distinct categories to characterize the approaches practitioners use to address the issues. As mentioned in Section \ref{subsubsec:FormalDataExtraction}, not all closed issues have an associated solution that can be identified. As a result, we identified a total of 427 solutions, which were used to address 45.2\% of the 944 issues, and categorized them into 10 categories as presented in Table~\ref{tab:solution_distribution}.

\subsubsection {Results} 
The results show that \textit{Optimize Workflow (OW)} is the most commonly utilized solution, accounting for 26.0\% of the total solutions, followed by \textit{Adjust Configuration (AC)}, \textit{Optimize Tool Integration (OTI)}, and \textit{Optimize Infrastructure (OI)}, accounting for 15.0\%, 11.9\%, and 11.0\%, respectively.

\begin{itemize}
    \item \textbf{\textsc{Optimize Workflow (OW)}} refers to improving workflow execution and task coordination in LLM-based MAS. It can be categorized into five types: (1) \textsc{Enhance Orchestration}: refers to adjusting strategies for coordination and task allocation among agents. (2) \textsc{Enhance State Management}: refers to adjusting strategies for managing execution states and intermediate data across agents. (3) \textsc{Fix Loop Issues}: refers to resolving infinite or unnecessary execution loops in workflow execution. (4) \textsc{Implement Parallel Execution}: refers to adjusting execution strategies to enable concurrent task processing across agents. (5) \textsc{Optimize Sequential Execution}: refers to optimizing execution strategies to ensure efficient workflow progression and data exchange between agents. For example, a user proposed ``\textit{adding an agent parameter to tool hooks and injecting data keys into the LLM context to reduce token usage and latency when transferring large datasets between agents}'' (Agno \#3712).

    \item \textbf{\textsc{Adjust Configuration (AC)}} refers to modifying system settings and parameters in LLM-based MAS. It can be categorized into four types: (1) \textsc{Refine Prompt}: refers to adjusting prompt design to guide agent behavior. (2) \textsc{Modify Runtime Parameter}: refers to adjusting execution parameters to control runtime behavior. For example, a user proposed ``\textit{adding a runtime parameter to execute only the latest message, preventing unintended code execution in multi-agent chats}'' (Autogen \#5283). (3) \textsc{Configure External Dependency}: refers to configuring external tools or libraries required for agents to perform tasks within LLM-based MAS. (4) \textsc{Update Environment Configuration}: refers to adjusting environment-level settings required for the compatibility and correct execution of LLM-based MAS.

    \begin{table*}[htbp]
\scriptsize
\caption{Solutions to Issues in Open-Source LLM-based Multi-Agent Projects}
\label{tab:solution_distribution}
\begin{tabular}{m{4.5cm}m{0.8cm}<{\centering}m{0.8cm}<{\centering}m{9cm}}
\hline
\textbf{Category} & \textbf{Count} & \textbf{\%} & \textbf{Example (Extracted Solution)} \\
\hline

\textit{Optimize Workflow (OW)} & 111 & 26.0\% & \textit{Replace the single agent with two agents, one for image parsing and another for retrieval and answering.} (Ragflow \#9451) \\
\hline

\textit{Adjust Configuration (AC)} & 64 & 15.0\% & \textit{Include custom message types during multi-agent system setup.} (AutoGen \#6795) \\
\hline

\textit{Optimize Tool Integration (OTI)} & 51 & 11.9\% & \textit{Move external tools from individual agents to the team to prevent delegation errors.} (Agno \#4759) \\
\hline

\textit{Optimize Infrastructure (OI)} & 47 & 11.0\% & \textit{Resolve database module loading compatibility to ensure infrastructure stability.} (Claude-flow \#893) \\
\hline

\textit{Improve Error Handling (IEH)} & 45 & 10.5\% & \textit{Use a function to end the conversation instead of relying on the AI to generate a termination word.} (AutoGen \#133) \\
\hline

\textit{Optimize Communication (OC)} & 33 & 7.7\% & \textit{Check message size before transmission and switch to streaming mode if exceeding the limit.} (AgentScope \#269) \\
\hline

\textit{Update Dependencies (UD)} & 31 & 7.3\% & \textit{Upgrade Docker images to newer compatible versions.} (Flowise \#4020) \\
\hline

\textit{Adjust Agent Initialization (AAI)} & 29 & 6.8\% & \textit{Before updating the agent during initialization, check for parameter changes.} (Agency-Swarm \#126)  \\
\hline

\textit{Improve Documentation (ID)} & 13 & 3.0\% & \textit{Add a documentation section with code examples showing how to use sub‑agents as tools.} (Adk-Python \#1608) \\
\hline

\textit{Improve Frontend (IF)} & 3 & 0.7\% & \textit{Add a user interface option for removing links between nodes in an LLM-based MAS.} (Flowise \#4494) \\
\hline
\end{tabular}
\end{table*}

    \item \textbf{\textsc{Optimize Tool Integration (OTI)}} refers to improving tool connectivity and execution behavior in LLM-based MAS. It can be categorized into two types: (1) \textsc{Fix Tool Execution}: refers to resolving tool execution problems caused by execution flow issues or errors in LLM-based MAS. (2) \textsc{Tool Integration Fix}: refers to resolving issues in tool registration, connectivity, or compatibility within LLM-based MAS. For example, a user suggested that ``\textit{connecting and closing MCP tools through the FastAPI application lifespan to resolve MCP connectivity failures}'' (Agno \#4958).

    \item \textbf{\textsc{Optimize Infrastructure (OI)}} refers to improving infrastructure components and resource management in LLM-based MAS. It can be categorized into two types: (1) \textsc{Database Compatibility}: refers to resolving incompatibilities in database backends and storage systems due to improper configuration in LLM-based MAS. (2) \textsc{Optimize Memory Management}: refers to improving memory initialization, synchronization, and persistence to ensure reliable memory handling across agents in LLM-based MAS. For example, a user resolved ``\textit{the unknown memory backend: undefined error by adding required configuration parameters and creating a default SQLite configuration with a 50MB cache and 30s synchronization interval}'' (Claude-flow \#354).

    \item \textbf{\textsc{Improve Error Handling (IEH)}} refers to the process of solving problems by improving error detection, exception management, and recovery behavior in LLM-based MAS. It can be categorized into four types: (1) \textsc{Fix Error Propagation}: refers to addressing situations in which an error generated by one agent or workflow step is propagated to subsequent agents or steps without being properly communicated, detected, or handled, potentially causing downstream execution failures. (2) \textsc{Enhance Fallback Mechanism}: refers to improving fallback strategies to maintain workflow execution when agents or tools fail in LLM-based MAS. (3) \textsc{Enhance Handoff Message Flow}: refers to improving message routing and conversation continuity during agent handoffs in LLM-based MAS. (4) \textsc{Enhance Exception Handling}: refers to improving exception management through retry logic, clearer error reporting, and recovery mechanisms to handle failures during LLM-based MAS execution. For example, a user proposed enhancing exception handling by ``\textit{adding retry logic, circuit breakers, health checks, and clearer API error messages to manage external service failures more effectively}'' (Claude-flow \#183).

    \item \textbf{\textsc{Optimize Communication (OC)}} refers to improving information exchange between agents in LLM-based MAS. It can be categorized into two types: (1) \textsc{Optimize Message Handling}: refers to improving message processing, including parsing, formatting, and response transformation within agents or communication modules in LLM-based MAS. (2) \textsc{Optimize Inter-Agent Communication}: refers to improving the exchange of messages between agents to ensure correct delivery, interpretation, and coordination across agents. For example, a user suggested that ``\textit{correcting the serialization of response data in the remote agent client can prevent agent-to-agent communication failures}'' (Adk-Python \#2921).

    \item \textbf{\textsc{Update Dependencies (UD)}} refers to solving dependency-related problems, including missing, mismatched, or incompatible libraries and package versions in LLM-based MAS. It can be categorized into two types: (1) \textsc{Add Missing Dependency}: refers to installing required libraries that are missing from the environment for LLM-based MAS. (2) \textsc{Fix Version Compatibility}: refers to resolving dependency version errors by aligning dependent library and package versions to ensure correct integration in LLM-based MAS. For example, a user resolved ``\textit{the issue by updating the outdated Pydantic dependency from version 2.5.2 to the latest version}'' (Agno \#2444).

    \item \textbf{\textsc{Adjust Agent Initialization (AAI)}} refers to adjusting agent setup and initialization in LLM-based MAS. It can be categorized into three types: (1) \textsc{Fix Import Path Resolution}: refers to resolving issues where agent modules or definitions cannot be correctly located or loaded in LLM-based MAS. (2) \textsc{Correct Agent Initialization}: refers to improving dynamic agent creation and registration to ensure that required agent types are correctly initialized and available in LLM-based MAS. For example, a user resolved ``\textit{the issue by replacing the incorrect knowledge\_sources parameter with the expected knowledge parameter when creating agents}'' (PraisonAI \#404). (3) \textsc{Configure Agent Integration}: refers to configuring the connections between agents and their required components, such as knowledge sources, models, or tools, during agent initialization in LLM-based MAS.
    
    \item \textbf{\textsc{Improve Documentation (ID)}} refers to improving the clarity and completeness of documentation in LLM-based MAS. It can be categorized into two types: (1) \textsc{Modify Documentation}: refers to modifying or updating existing documentation to better clarify system behavior or usage instructions in LLM-based MAS. (2) \textsc{Add Documentation}: refers to adding missing guidance, examples, or workflow descriptions for system features in LLM-based MAS. For example, a user provided ``\textit{documentation and examples for AgentTool usage in multi-agent workflows}'' (Adk-Python \#1608).

    \item \textbf{\textsc{Improve Frontend (IF)}} refers to improving the user interface and output presentation of LLM-based MAS. It can be categorized into two types: (1) \textsc{Enhance UI Interaction}: refers to improving how users interact with LLM-based MAS through interface controls, input mechanisms, and interactive features. (2) \textsc{Enhance UI Display}: refers to improving the presentation, updating, and streaming of agent outputs in LLM-based MAS. For example, a user proposed ``\textit{displaying agent names in chat messages to clarify which agent is communicating}'' (AutoGen \#5609).
\end{itemize}

\subsubsection{Solutions to Issues Mapping}\label{subsec:SolutionstoIssuesMapping}
Table \ref{tab:issues-solutions-mapping} illustrates the mapping of \textit{Issue} categories to \textit{Solution} categories, along with their distribution, using abbreviations to represent each category of solution. For example, “OTI” represents \textit{Optimize Tool Integration}. The full names of all solution categories are provided in the note of Table \ref{tab:issues-solutions-mapping}.

\begin{table*}[h]
\scriptsize
\caption{ Mapping between issue categories (vertical) and solution categories (horizontal)}
\label{tab:issues-solutions-mapping}
\begin{adjustbox}{center}
\begin{tabular}{ccccccccccc}
                                                    & \multicolumn{1}{c}{\textbf{OW}} & \multicolumn{1}{c}{\textbf{AC}} & \multicolumn{1}{c}{\textbf{OTI}} & \multicolumn{1}{c}{\textbf{OC}} 
                                                    & \multicolumn{1}{c}{\textbf{UD}} & \multicolumn{1}{c}{\textbf{AAI}} & \multicolumn{1}{c}{\textbf{IEH}} & \multicolumn{1}{c}{\textbf{OI}} 
                                                    & \multicolumn{1}{c}{\textbf{ID}} & \multicolumn{1}{c}{\textbf{IF}} \\
\rowcolor[HTML]{acddde} \textbf{Orchestration \& Execution Issue} 
                                                    & \cellcolor[HTML]{5da6d6}38      & \cellcolor[HTML]{cce8f7}5        & \cellcolor[HTML]{aad4ed}6        & \cellcolor[HTML]{e9f7ff}2 
                                                    & \cellcolor[HTML]{e9f7ff}1       & \cellcolor[HTML]{e9f7ff}2       & \cellcolor[HTML]{aad4ed}14        & \cellcolor[HTML]{e9f7ff}10 
                                                    & \cellcolor[HTML]{e9f7ff}1       & \cellcolor[HTML]{f4f8fb}0 \\
\rowcolor[HTML]{e2f8dc} \textbf{Memory \& State Management Issue} 
                                                    & \cellcolor[HTML]{aad4ed}17      & \cellcolor[HTML]{cce8f7}5        & \cellcolor[HTML]{e9f7ff}2        & \cellcolor[HTML]{cce8f7}4 
                                                    & \cellcolor[HTML]{e9f7ff}2       & \cellcolor[HTML]{e9f7ff}2       & \cellcolor[HTML]{e9f7ff}5        & \cellcolor[HTML]{5da6d6}23 
                                                    & \cellcolor[HTML]{f4f8fb}0       & \cellcolor[HTML]{f4f8fb}0 \\
\rowcolor[HTML]{acddde} \textbf{Feature Request} 
                                                    & \cellcolor[HTML]{aad4ed}10      & \cellcolor[HTML]{cce8f7}7       & \cellcolor[HTML]{cce8f7}7        & \cellcolor[HTML]{aad4ed}12 
                                                    & \cellcolor[HTML]{e9f7ff}0       & \cellcolor[HTML]{f4f8fb}0       & \cellcolor[HTML]{cce8f7}4        & \cellcolor[HTML]{e9f7ff}3 
                                                    & \cellcolor[HTML]{e9f7ff}1       & \cellcolor[HTML]{f4f8fb}0 \\
\rowcolor[HTML]{e2f8dc} \textbf{Implementation Issue} 
                                                    & \cellcolor[HTML]{aad4ed}18      & \cellcolor[HTML]{aad4ed}16       & \cellcolor[HTML]{aad4ed}8        & \cellcolor[HTML]{cce8f7}4 
                                                    & \cellcolor[HTML]{cce8f7}4       & \cellcolor[HTML]{aad4ed}10        & \cellcolor[HTML]{cce8f7}6        & \cellcolor[HTML]{e9f7ff}2 
                                                    & \cellcolor[HTML]{f4f8fb}0       & \cellcolor[HTML]{e9f7ff}1 \\
\rowcolor[HTML]{acddde} \textbf{Tool Issue} 
                                                    & \cellcolor[HTML]{aad4ed}10      & \cellcolor[HTML]{cce8f7}4        & \cellcolor[HTML]{aad4ed}12       & \cellcolor[HTML]{e9f7ff}1 
                                                    & \cellcolor[HTML]{cce8f7}7       & \cellcolor[HTML]{e9f7ff}2        & \cellcolor[HTML]{e9f7ff}2        & \cellcolor[HTML]{e9f7ff}1 
                                                    & \cellcolor[HTML]{f4f8fb}0       & \cellcolor[HTML]{e9f7ff}1 \\
\rowcolor[HTML]{e2f8dc} \textbf{Integration Issue} 
                                                    & \cellcolor[HTML]{e9f7ff}2       & \cellcolor[HTML]{aad4ed}7        & \cellcolor[HTML]{aad4ed}7       & \cellcolor[HTML]{e9f7ff}1 
                                                    & \cellcolor[HTML]{cce8f7}6       & \cellcolor[HTML]{cce8f7}4        & \cellcolor[HTML]{e9f7ff}2        & \cellcolor[HTML]{e9f7ff}2 
                                                    & \cellcolor[HTML]{e9f7ff}1       & \cellcolor[HTML]{f4f8fb}0 \\
\rowcolor[HTML]{acddde} \textbf{Role Issue} 
                                                    & \cellcolor[HTML]{cce8f7}3       & \cellcolor[HTML]{cce8f7}5        & \cellcolor[HTML]{e9f7ff}1        & \cellcolor[HTML]{e9f7ff}0 
                                                    & \cellcolor[HTML]{f4f8fb}0       & \cellcolor[HTML]{cce8f7}5        & \cellcolor[HTML]{e9f7ff}2        & \cellcolor[HTML]{e9f7ff}1 
                                                    & \cellcolor[HTML]{f4f8fb}0       & \cellcolor[HTML]{f4f8fb}0 \\
\rowcolor[HTML]{e2f8dc} \textbf{Environment \& Configuration Issue} 
                                                    & \cellcolor[HTML]{f4f8fb}0       & \cellcolor[HTML]{aad4ed}5        & \cellcolor[HTML]{e9f7ff}1        & \cellcolor[HTML]{e9f7ff}1 
                                                    & \cellcolor[HTML]{aad4ed}8       & \cellcolor[HTML]{e9f7ff}1        & \cellcolor[HTML]{f4f8fb}0        & \cellcolor[HTML]{e9f7ff}1 
                                                    & \cellcolor[HTML]{e9f7ff}1       & \cellcolor[HTML]{f4f8fb}0 \\
\rowcolor[HTML]{acddde} \textbf{Output Issue} 
                                                    & \cellcolor[HTML]{e9f7ff}1       & \cellcolor[HTML]{f4f8fb}0        & \cellcolor[HTML]{e9f7ff}1        & \cellcolor[HTML]{e9f7ff}0 
                                                    & \cellcolor[HTML]{f4f8fb}0       & \cellcolor[HTML]{e9f7ff}1        & \cellcolor[HTML]{e9f7ff}2        & \cellcolor[HTML]{e9f7ff}1 
                                                    & \cellcolor[HTML]{f4f8fb}0       & \cellcolor[HTML]{e9f7ff}1 \\
\rowcolor[HTML]{e2f8dc} \textbf{Documentation Issue} 
                                                    & \cellcolor[HTML]{e9f7ff}2       & \cellcolor[HTML]{e9f7ff}2        & \cellcolor[HTML]{e9f7ff}2        & \cellcolor[HTML]{f4f8fb}0 
                                                    & \cellcolor[HTML]{e9f7ff}1       & \cellcolor[HTML]{e9f7ff}1        & \cellcolor[HTML]{e9f7ff}2        & \cellcolor[HTML]{e9f7ff}1 
                                                    & \cellcolor[HTML]{aad4ed}9       & \cellcolor[HTML]{f4f8fb}0 \\
\rowcolor[HTML]{acddde} \textbf{Communication Issue} 
                                                    & \cellcolor[HTML]{e9f7ff}3       & \cellcolor[HTML]{e9f7ff}2        & \cellcolor[HTML]{f4f8fb}0        & \cellcolor[HTML]{aad4ed}4 
                                                    & \cellcolor[HTML]{f4f8fb}0       & \cellcolor[HTML]{f4f8fb}0        & \cellcolor[HTML]{f4f8fb}0        & \cellcolor[HTML]{f4f8fb}0 
                                                    & \cellcolor[HTML]{f4f8fb}0       & \cellcolor[HTML]{f4f8fb}0 \\
\rowcolor[HTML]{e2f8dc} \textbf{Performance Issue} 
                                                    & \cellcolor[HTML]{e9f7ff}1       & \cellcolor[HTML]{e9f7ff}2        & \cellcolor[HTML]{e9f7ff}1        & \cellcolor[HTML]{f4f8fb}0 
                                                    & \cellcolor[HTML]{f4f8fb}0       & \cellcolor[HTML]{e9f7ff}1        & \cellcolor[HTML]{f4f8fb}0        & \cellcolor[HTML]{f4f8fb}0 
                                                    & \cellcolor[HTML]{f4f8fb}0       & \cellcolor[HTML]{f4f8fb}0 \\
\rowcolor[HTML]{acddde} \textbf{Model Issue} 
                                                    & \cellcolor[HTML]{e9f7ff}1       & \cellcolor[HTML]{e9f7ff}1        & \cellcolor[HTML]{e9f7ff}2        & \cellcolor[HTML]{e9f7ff}1 
                                                    & \cellcolor[HTML]{f4f8fb}0       & \cellcolor[HTML]{f4f8fb}0        & \cellcolor[HTML]{f4f8fb}0        & \cellcolor[HTML]{e9f7ff}1 
                                                    & \cellcolor[HTML]{f4f8fb}0       & \cellcolor[HTML]{f4f8fb}0 \\
\rowcolor[HTML]{e2f8dc} \textbf{Compatibility Issue} 
                                                    & \cellcolor[HTML]{e9f7ff}1       & \cellcolor[HTML]{e9f7ff}1        & \cellcolor[HTML]{e9f7ff}1        & \cellcolor[HTML]{e9f7ff}2 
                                                    & \cellcolor[HTML]{e9f7ff}2       & \cellcolor[HTML]{f4f8fb}0        & \cellcolor[HTML]{e9f7ff}1        & \cellcolor[HTML]{f4f8fb}0 
                                                    & \cellcolor[HTML]{f4f8fb}0       & \cellcolor[HTML]{f4f8fb}0 \\
\rowcolor[HTML]{acddde} \textbf{Error Handling Issue} 
                                                    & \cellcolor[HTML]{e9f7ff}1       & \cellcolor[HTML]{e9f7ff}1        & \cellcolor[HTML]{f4f8fb}0        & \cellcolor[HTML]{f4f8fb}0 
                                                    & \cellcolor[HTML]{f4f8fb}0       & \cellcolor[HTML]{f4f8fb}0        & \cellcolor[HTML]{aad4ed}5        & \cellcolor[HTML]{e9f7ff}1 
                                                    & \cellcolor[HTML]{f4f8fb}0       & \cellcolor[HTML]{f4f8fb}0 \\
\rowcolor[HTML]{e2f8dc} \textbf{Task Management Issue} 
                                                    & \cellcolor[HTML]{cce8f7}3       & \cellcolor[HTML]{e9f7ff}1        & \cellcolor[HTML]{f4f8fb}0        & \cellcolor[HTML]{f4f8fb}0 
                                                    & \cellcolor[HTML]{f4f8fb}0       & \cellcolor[HTML]{f4f8fb}0        & \cellcolor[HTML]{f4f8fb}0        & \cellcolor[HTML]{f4f8fb}0 
                                                    & \cellcolor[HTML]{f4f8fb}0       & \cellcolor[HTML]{f4f8fb}0 \\
\rowcolor[HTML]{acddde} \textbf{Security Issue} 
                                                    & \cellcolor[HTML]{f4f8fb}0       & \cellcolor[HTML]{f4f8fb}0        & \cellcolor[HTML]{f4f8fb}0        & \cellcolor[HTML]{e9f7ff}1 
                                                    & \cellcolor[HTML]{f4f8fb}0       & \cellcolor[HTML]{f4f8fb}0        & \cellcolor[HTML]{f4f8fb}0        & \cellcolor[HTML]{f4f8fb}0 
                                                    & \cellcolor[HTML]{f4f8fb}0       & \cellcolor[HTML]{f4f8fb}0 \\

\end{tabular}
\end{adjustbox}
\begin{minipage}{\textwidth}
\vspace{0.1cm}
\vspace{0.1cm}
\noindent\textbf{Full names of every solution category abbreviation:}
OW: \textit{Optimize Workflow}; AC: \textit{Adjust Configuration}; OTI: \textit{Optimize Tool Integration}; OC: \textit{Optimize Communication}; UD: \textit{Update Dependencies}; AAI: \textit{Adjust Agent Initialization}; IEH: \textit{Improve Error Handling}; OI: \textit{Optimize Infrastructure}; ID: \textit{Improve Documentation}; IF: \textit{Improve Frontend}.
\end{minipage}
\end{table*}

For \textit{Documentation Issue}, 76.9\% of cases are well addressed, reaching the highest resolution rate among all issue categories. \textit{Improve Documentation (ID)} (9) is identified as the most frequent solution for resolving \textit{Documentation Issue}.

For \textit{Output Issue}, 24.1\% of cases are reported as resolved, reaching the lowest resolution rate among all issue categories, and \textit{Output Issue} is mostly solved by \textit{Improve Error Handling (IEH)} (2).

For \textit{Orchestration \& Execution Issue}, which is the largest issue category, 37.9\% of cases have effective solutions. \textit{Orchestration \& Execution Issue} is mostly addressed by \textit{Optimize Workflow (OW)} (38).

For \textit{Implementation Issue}, 69.0\% of cases are well addressed. The most frequently used solution for \textit{Implementation Issue} is \textit{Optimize Workflow (OW)} (18).

For \textit{Compatibility Issue}, solutions are identified in 61.5\% of cases. \textit{Update Dependencies (UD)} (2) and \textit{Optimize Communication (OC)} (2) are the most frequently employed solutions to resolve \textit{Compatibility Issue}.

For \textit{Error Handling Issue}, 66.7\% of cases are well resolved. The most frequently employed solution for \textit{Error Handling Issue} is \textit{Improve Error Handling (IEH)} (5).

For \textit{Environment \& Configuration Issue}, 54.5\% of cases have effective solutions. \textit{Update Dependencies (UD)} (8) is identified as the most frequently used solution to address the \textit{Environment \& Configuration Issue}.

For \textit{Integration Issue}, solutions are identified in 50.8\% of cases. The most frequently employed solutions for \textit{Integration Issue} are \textit{Adjust Configuration (AC)} (7) and \textit{Optimize Tool Integration (OTI)} (7).

For \textit{Role Issue}, 47.2\% of cases are effectively addressed. The most frequently employed solutions to address \textit{Role Issues} are \textit{Adjust Agent Initialization (AAI)} (5) and \textit{Adjust Configuration (AC)} (5).

For \textit{Memory \& State Management Issue}, 43.2\% of cases are addressed by practitioners. The most frequently employed solution for \textit{Memory \& State Management Issue} is \textit{Optimize Infrastructure (OI)} (23).

For \textit{Tool Issue}, solutions are identified in 45.9\% of cases. \textit{Optimize Tool Integration (OTI)} (12) is the most commonly applied solution to resolve \textit{Tool Issue}.

For \textit{Model Issue}, 40.0\% of cases are resolved. The most frequently used solution to address \textit{Model Issue} is \textit{Optimize Tool Integration (OTI)} (2).

For \textit{Task Management Issue}, 36.4\% of cases are resolved by practitioners. \textit{Optimize Workflow (OW)} (3) is identified as the most commonly applied solution to address \textit{Task Management Issue}.

For \textit{Feature Request}, 35.2\% of cases are well addressed by practitioners. The most commonly employed solution for \textit{Feature Requests} is \textit{Optimize Communication (OC)} (12).

For \textit{Communication Issue}, 36.0\% of cases are well resolved by practitioners. \textit{Optimize Communication (OC)} (4) is identified as the most frequently employed solution for \textit{Communication Issue}.

For \textit{Security Issue}, 33.3\% of cases are resolved. The most frequently used solution to address \textit{Security Issue} is \textit{Optimize Communication (OC)} (1).

For \textit{Performance Issue}, 26.3\% of cases are well resolved by practitioners. \textit{Adjust Configuration (AC)} (2) is identified as the most frequently employed solution for \textit{Performance Issue}.

\subsubsection{Interpretation}\label{sec:interpretation} 
In this section, we analyze the frequency and distribution of different solutions across various issue categories in LLM-based MAS. We further map these solutions to specific issue types to identify the dominant solutions used to address different issues.

\textbf{\emph{Frequency of the solutions}}: \textit{Optimize Workflow (OW)}, \textit{Adjust Configuration (AC)}, \textit{Optimize Tool Integration (OTI)}, and \textit{Optimize Infrastructure (OI)} are the most frequently used solutions in LLM-based MAS due to their direct impact on execution flow, configuration, tool usage, and infrastructure reliability. \textit{Optimize Workflow (OW)} is widely applied because MAS depend on proper orchestration, state sharing, and task coordination to ensure smooth execution. \textit{Adjust Configuration (AC)} is also a common solution since incorrect prompts, runtime settings, or environment configurations often lead to system errors, making parameter adjustment a frequent fix strategy in LLM-based MAS. \textit{Optimize Tool Integration (OTI)} is another major category of solutions, as LLM-based MAS rely heavily on external tools and APIs, where integration or execution issues can easily break workflows between agents. \textit{Optimize Communication (OC)} is a similarly important solution category because reliable inter-agent communication is essential for correct message exchange and task completion in LLM-based MAS. Although \textit{Update Dependencies (UD)}, \textit{Adjust Agent Initialization (AAI)}, \textit{Improve Error Handling (IEH)}, \textit{Improve Documentation (ID)}, and \textit{Improve Frontend (IF)} are employed less frequently, they remain important for MAS stability. For example, \textit{Optimize Infrastructure (OI)} shows that practitioners resolve system failures by improving memory and database infrastructure to ensure stable execution and data persistence in LLM-based MAS.

\textbf{\emph{Mapping of solutions to issues}}: \textit{Optimize Workflow (OW)} is the most dominant solution category, accounting for 26.0\% of all identified solutions. For \textit{Orchestration \& Execution Issue}, which is the largest issue category, \textit{Optimize Workflow (OW)} addresses 48.1\% of cases, reflecting the critical role of workflow optimization in resolving agent coordination and execution failures. Similarly, \textit{Adjust Configuration (AC)} is the second most common solution (15.0\%), and it accounts for 21.9\% of solutions for \textit{Integration Issue}. \textit{Optimize Tool Integration (OTI)} accounts for 11.9\% of solutions and is the most frequent solution for \textit{Tool Issues}, accounting for 30.0\% of cases, reflecting the reliance of MAS on external tools and the need for robust tool integration mechanisms. \textit{Optimize Infrastructure (OI)} is the fourth most common solution, accounting for 11.0\% of all identified solutions, and is predominantly applied to address \textit{Memory \& State Management Issues}. \textit{Documentation Issue}  has the highest resolution rate of 76.0\%, while \textit{Output Issue} has the lowest resolution rate 24.1\%. Overall, the predominance of \textit{Optimize Workflow (OW)}, \textit{Adjust Configuration (AC)}, \textit{Optimize Tool Integration (OTI)}, and \textit{Optimize Infrastructure (OI)}  across multiple issue categories indicates that practitioners primarily focus on workflow execution, configuration management, and tool integration when resolving issues in LLM-based MAS.

\section{Implications}\label{sec:Implications}
\label{sec:implication}
In this section, we discuss the implications of the study results presented in Section~\ref{sec:ResultsandInterpretation} for users, developers, and researchers of open-source LLM-based MAS.
\subsection{Implications for Users of LLM-based MAS}

\begin{tcolorbox} [sharp corners, boxrule=0.1mm,]
\small
\textbf{Implication 1}. Users can address the \textit{Orchestration \& Execution Issue} by designing explicit prompt-based handoff mechanisms and implementing appropriate execution controls.
\end{tcolorbox}

\textit{Orchestration \& Execution Issue} (22.0\%) ranks first among all 17 issue categories, and within this category, \textit{Coordination \& Handoff Issue} is the dominant type, indicating that users frequently face difficulties in coordinating task transfers and managing execution flow in LLM-based MAS. A practitioner reported that ``\textit{the sub-agent could not correctly transfer control back to the root agent when calling an MCP tool}'' (Adk-python \#1039), suggesting that unclear handoff specifications can lead to agents coordination failures. Several practices can improve coordination among agents and enhance the reliability of task execution in LLM-based MAS. For example, users can define explicit handoff prompts or utilize built-in agent handoff mechanisms provided by frameworks such as OpenAI Agents SDK \citep{openai2025agents} to clearly specify transfer conditions and task responsibilities between agents. Furthermore, users can configure framework-level execution controls, such as retry limits, timeout settings, and logging mechanisms, to reduce infinite loops, premature termination, and unexpected execution behaviors in LLM-based MAS. In addition, many LLM-based multi-agent frameworks provide orchestration and monitoring capabilities that support workflow design, inspection, and debugging, enabling early detection of coordination issues and improving collaboration reliability between agents \citep{dibia2024autogenstudio}. For example, cite{cemri2025multi} conducted a comprehensive empirical study on multi-agent failure modes and found that inter-agent misalignment, including communication breakdowns, context loss during handoffs, and conflicting outputs, accounts for 36.9\% of failures across seven MAS frameworks. Consequently, users should prioritize explicit handoff design and systematic execution monitoring to address \textit{Orchestration \& Execution Issue}, the most common issue category in LLM-based MAS.

\begin{tcolorbox} [sharp corners, boxrule=0.1mm,]
\small
\textbf{Implication 2}. Users can leverage LLM-based multi-agent development frameworks for software development tasks while remaining mindful of the potential issues and limitations of LLMs.
\end{tcolorbox}

One-third of the open-source LLM-based MAS investigated in this study are frameworks for developing software applications by leveraging multiple collaborating LLM agents (see Table \ref{tab:MAS_domain_categorization}). These frameworks lower the barrier for users to build applications using LLM-based MAS. For example, \texttt{AutoGen} is an open-source framework that enables users to build LLM applications using multiple agents that can communicate with each other to accomplish tasks \citep{wu2024autogen}. It also allows users to configure agents with different commercial and open-source LLMs, according to their application requirements, enabling users to select a model that best fits their needs. While such frameworks provide useful capabilities for software development, our findings show that LLM-based MAS also face several problems. Specifically, \textit{Orchestration \& Execution Issue}, \textit{Memory \& State Management Issue}, \textit{Tool Issue}, and \textit{Integration Issue} account for 22.0\%, 14.7\%, 9.2\%, and 6.7\% of the issues, respectively. These findings suggest that using open-source LLM-based MAS for software development may introduce various challenges. Users should therefore monitor agent workflows, manage agent memory and state, verify tool usage, and configure explicit agent communication mechanisms when deploying and using LLM-based MAS.

\subsection{Implications for Developers of LLM-based MAS}

\begin{tcolorbox} [sharp corners, boxrule=0.1mm,]
\small
\textbf{Implication 3}. Developers of open-source LLM-based MAS should possess a foundational understanding of multi-agent system design principles.
\end{tcolorbox}

According to our study results, the following knowledge is required to develop and use open-source LLM-based MAS: (1) Developers need knowledge of designing and managing agent workflows and orchestration in LLM-based MAS. Our findings show that \textit{Workflow Problem (WP)} accounts for 29.4\% and ranks first among all identified causes. This indicates that developers need solid knowledge of agent coordination, handoff design, and execution flow management when developing LLM-based MAS. For example, a contributor reported that ``\textit{after invoking transfer\_to\_agent, the target agent did not respond and produced no output due to incomplete transfer of task-related information}'' (Adk-python \#2073), highlighting the importance of proper handoff design while developing LLM-based MAS. (2) Developers need to understand tool integration and execution in LLM-based MAS. According to our study results, \textit{Tool Integration Problem (TIP)} accounts for 11.7\% of all causes. This indicates that developers require knowledge of tool registration, schema validation, and concurrent tool access management when developing LLM-based MAS. For example, a developer reported that a ``\textit{tool deadlock occurred because the swarm could not access tools already locked by an active Claude Code session}'' (Claude-flow \#59), highlighting that knowledge of proper tool lifecycle management is essential when developing LLM-based MAS. (3) Developers need to understand memory management and state persistence in LLM-based MAS. \textit{Memory Problem (MP)} accounts for 11.1\% of all causes. For example, a developer explained that ``\textit{external memory failed to store new memories in Mem0 because the agent ID exceeded the allowed length}'' (CrewAI \#2985), indicating that knowledge of session persistence and state consistency is essential for developing reliable LLM-based MAS.

\begin{tcolorbox} [sharp corners, boxrule=0.1mm,]
\small
\textbf{Implication 4}. Developers can enhance the reliability of tool integration through schema validation and robust execution controls.
\end{tcolorbox}

\textit{Tool Integration Problem (TIP)} accounts for 11.7\% of all identified causes, with \textit{Tool Execution Problem} and \textit{MCP Integration Problem} being the most frequently observed types. A typical failure occurs when tools encounter deadlocks, resource contention, or schema mismatches at runtime in LLM-based MAS. For example, a practitioner reported that ``\textit{MCP tool calls failed during asynchronous GroupChat execution because the event loop was already running}'' (Ag2 \#1763), while another reported that ``\textit{VertexAI Search Tool failed when combined with a sub-agent because multiple tool types were not supported together}'' (Adk-Python \#2822). To avoid such failures, developers should provide standardized tool validation APIs that automatically check input/output schemas when tools are configured, enforce timeouts at the tool level, and check tool availability before execution in LLM-based MAS. Additionally, since concurrent tool access by multiple agents can cause session locks, connection failures, and timeouts, MAS frameworks should provide mechanisms for managing tool connections, handling tool connection failures, and retrying failed tool calls. One practitioner reported that ``\textit{managing tool connections through framework-provided lifecycle hooks (e.g., FastAPI lifespan in Agno) prevented MCP connection failures}'' (Agno \#4958), highlighting the importance of proper tool lifecycle management for reliable MCP integration in LLM-based MAS.

\subsection{Implications for Researchers of LLM-based MAS}

\begin{tcolorbox} [sharp corners, boxrule=0.1mm,]
\small
\textbf{Implication 5}. Researchers may consider exploring orchestration frameworks to improve agent coordination and execution reliability in LLM-based MAS.
\end{tcolorbox}

\textit{Orchestration \& Execution Issue}, accounting for 22.0\% of all reported issues, represents the most critical challenge in LLM-based MAS. Within this category, \textit{Coordination \& Handoff Issue} and \textit{Control Flow Error} are the most frequent types, highlighting that failures often originate from coordination deficiencies and execution management challenges among agents. This finding is further supported by the root cause analysis, which identifies \textit{Workflow Problem (WP)} as the most significant underlying factor, accounting for 29.4\% of all reported causes. Moreover, \textit{Workflow Problem (WP)} accounts for 57.0\% of the causes associated with \textit{Orchestration \& Execution Issue}. This indicates that deficiencies in multi-agent workflow design, execution flow among agents, and inter-agent task coordination are major factors contributing to system-level failures in LLM-based MAS. We encourage researchers to shift their focus from enhancing individual agents reasoning capabilities toward developing robust orchestration frameworks that ensure reliable handoffs between agents, correct execution order across the agents, and appropriate termination conditions in LLM-based MAS. Specifically, future work could investigate the formal verification of agent workflows and the development of runtime monitoring mechanisms capable of detecting and correcting coordination anomalies, such as infinite loops and premature termination, before they propagate throughout the system.

\begin{tcolorbox} [sharp corners, boxrule=0.1mm,]
\small
\textbf{Implication 6}. Researchers could investigate methods for improving the reproducibility of issues in LLM-based MAS.
\end{tcolorbox}

Given the complex and dynamic behavior of LLM-based MAS, 27 of the 128 unsolved issues involve difficulties in reproducing the reported problems. Without accurately reproducing the problem within LLM-based MAS, it can be difficult for practitioners to identify root causes of the issues, thereby hindering the resolution of the issues. For example, practitioners reported that the ``\textit{transfer\_task\_to\_member error occasionally occurs in some random cases when using the aprint\_response method}'' (Agno \#3095), with multiple users experiencing different behaviors across versions. Identifying the root cause, in this case, a shared model instance issue, requires substantial investigation, highlighting the challenges of reproducing issues in LLM-based MAS and complicating their resolution. Therefore, future research could explore automated failure-reproduction techniques, controlled execution environments, and systematic test-case generation to improve the reproducibility of failures in LLM-based MAS.

\begin{tcolorbox} [sharp corners, boxrule=0.1mm,]
\small
\textbf{Implication 7}. Researchers could investigate reliable memory mechanisms for LLM-based MAS.
\end{tcolorbox}

\textit{Memory \& State Management Issue} represents the second most frequent challenge in LLM-based MAS, accounting for 14.7\% of all reported issues. Within this category, \textit{Context Loss} and \textit{State Inconsistency} are the most common types, indicating limitations in current memory mechanisms in LLM-based MAS. This finding is further supported by the root cause analysis, which identifies \textit{Memory Problem (MP)} as the third most significant underlying cause among all identified causes, accounting for 11.1\% of all reported causes. \textit{Memory \& State Management Issue} accounts for 51.1\% of the issues caused by \textit{Memory Problem (MP)}. This indicates that deficiencies in context propagation between agents, state synchronization across agents, and memory retrieval for agent interactions are major contributors to system failures in LLM-based MAS. We recommend that future research focus on developing reliable memory management mechanisms that ensure (1) consistent context propagation across agents during handoffs, (2) reliable state synchronization during parallel agent execution, and (3) effective context compression to prevent context overflow while preserving information essential for agent reasoning.

\section{Threats to Validity}\label{sec:ThreatstoValidity}
\textbf{Construct validity}: In our study, data collection was conducted automatically using GitHub APIs, while data filtering and labeling involved both automated and manual processes. Since several phases, including manual filtering, open coding, constant comparison, and data extraction, were performed manually, the study may be subject to personal bias during the classification, labeling, and extraction of issues, causes, and solutions in LLM-based MAS. To reduce this risk and improve the objectivity and reliability of the results, pilot studies were conducted using both automated and manual approaches before each formal phase to evaluate the effectiveness of methods, prompts, and criteria across different researchers, thereby improving construct validity. Following each pilot and formal study, we first performed automated filtering to obtain initial results. The first author then conducted a manual analysis and discussed the findings with the second and third authors to reach a consensus on the results related to the RQs. Since open coding and constant comparison were mainly carried out by the first author, there is still a risk of subjective interpretation affecting the results. The first author's six years of software development experience, along with two years of experience working with LLM-based MASs, helped mitigate this risk, although it could not be eliminated entirely. In addition, future work may adopt more advanced automated or semi-automated techniques to further reduce bias in data filtering, labeling, and extraction.

\textbf{External validity}: In our study, the main threat to external validity comes from the selection of data sources. To address this, we used the “Issues” pages of GitHub projects as the primary data source. GitHub is the largest platform for hosting open-source projects. Each project includes an “Issues” section. Practitioners use it to report bugs, errors, and development difficulties. It also allows discussion between users and developers. To extract more reliable information on causes and solutions, we focused on closed issues. These usually contain resolved discussions and more complete resolution details of issues. However, even with these selection choices, we acknowledge that the dataset may not fully represent all practices in the development and use of LLM-based MAS. Some relevant information may reside outside GitHub or in undocumented settings, which could be explored as potential sources in future work.

\textbf{Reliability}: To address potential uncertainties arising from the research methodology, we took several steps to improve the reliability of this study. As described in Section~\ref{subsubsec:PilotDataFiltering}, we first conducted pilot studies to check the agreement between the first and third authors on data labeling before starting the formal study. In this pilot phase, Cohen’s Kappa value was 0.82, which indicated a good level of agreement. During the data labeling, extraction, and analysis stages, we regularly discussed cases where there were differences in interpretation. These discussions helped us resolve disagreements and keep the results consistent across different stages of the study. This also improved the overall stability of the findings. In addition, the dataset used in this study has been made publicly available \citep{replicationPackage}, so that other researchers can replicate the study and validate the results independently.

\section{Related Work} \label{sec:relatedwork}
The existing literature is organized into three categories. We first review empirical studies using software repository mining to investigate practitioner-reported issues, causes, and solutions in software systems and LLM-based projects in Section~\ref{subsec:MSRStudy}. We then discuss LLM-based MAS for software engineering in Section~\ref{subsec:MASinSE}, followed by studies exploring challenges, failure patterns, and limitations of LLM-based MAS in Section~\ref{subsec:MSRinMAS}. Finally, we summarize the research gaps motivating this study in Section~\ref{subsec:ConclusiveSummary}.

\subsection{Mining Software Repositories for Exploring Issues, Causes, and Solutions}\label{subsec:MSRStudy}
Mining software repositories has become a widely used approach for understanding practitioner-reported issues, their underlying causes, and corresponding solutions. \cite{waseem2021microservices} conducted an empirical study to understand issues in open-source microservice systems. They manually analyzed 1,345 closed GitHub issues from five projects and developed a taxonomy of 17 issue categories. Their results show that \textit{Technical Debt}, \textit{Build}, and \textit{Security} are the most frequently reported categories. They later extended this work by analyzing 2,641 GitHub issues from 15 open-source microservice systems. They also conducted 15 practitioner interviews and surveyed 150 practitioners. Based on these findings, they developed comprehensive taxonomies of issues, root causes, and solutions in microservice systems. \cite{zhao2023performance} conducted a large-scale empirical study of 570 performance issues from 13 open-source projects. They identified eight common root causes and corresponding solutions. Their findings show that 27\% of the issues require design-level optimization, highlighting the importance of architectural improvements for addressing performance issues. With the growing adoption of AI and LLMs in software engineering, researchers have extended repository mining to investigate practitioner-reported issues in LLM-based projects. \cite{cai2025demystifying} analyzed 994 GitHub issues from 15 open-source LLM projects. They developed taxonomies of issue categories, root causes, and solutions. Their results show that \textit{Model Issue} is the most common issue category, while \textit{Model Problem}, \textit{Configuration Problem}, and \textit{Connection Problem} are the most frequent root causes. \textit{Model Optimization} is the predominant solution. \cite{zhou2025copilot} investigated GitHub Copilot, an LLM-based AI programming assistant, by analyzing 473 GitHub issues, 706 GitHub discussions, and 142 Stack Overflow posts. Their findings show that \textit{Operation Issue} and \textit{Compatibility Issue} are the most common problem categories. \textit{Copilot Internal Error}, \textit{Network Connection Error}, and \textit{IDE Compatibility Issue} are the most frequent causes, while \textit{Bug Fixed by Copilot}, \textit{Modify Configuration}, and \textit{Use Suitable Version} are the predominant solutions. Collectively, these studies establish a foundation for understanding practitioner-reported issues, their root causes, and solutions through repository mining across different software domains.

\subsection{LLM-based Multi-Agent Systems for Software Engineering}\label{subsec:MASinSE}
Recent advances in LLMs have led to the emergence of LLM-based MAS, where multiple specialized agents collaborate to automate software engineering tasks. \cite{qian2024chatdev} introduced ChatDev, a chat-powered software development framework that assigns specialized roles (e.g., CEO, CTO, Programmer, Reviewer, and Tester) to LLM-based agents. By using a chat chain to decompose tasks and support sequential interactions among agents, ChatDev enables collaborative software development and achieves an 88\% executability score, reflecting the proportion of software projects generated by ChatDev that successfully compile and execute. These results demonstrate the effectiveness of LLM-based MAS for software development. Likewise, \cite{hong2024metagpt} introduced MetaGPT, a meta-programming framework that incorporates Standard Operating Procedures (SOPs) into multi-agent collaboration. By assigning specialized roles to agents and enforcing structured workflows, MetaGPT enables agents to collaboratively perform software engineering tasks through role-based coordination. Similarly, \cite{wu2024autogen} introduced AutoGen, a generalized multi-agent conversation framework for developing LLM-based applications. AutoGen enables agents to perform tasks using LLMs, human feedback, and external tools, while supporting flexible communication patterns for designing complex multi-agent workflows. AutoGen achieves 69.48\% accuracy on the MATH dataset, a standard benchmark comprising 12,500 high-school competition math problems, substantially outperforming vanilla GPT-4. In addition, \cite{nguyen2025agilecoder} introduced AgileCoder, a multi-agent software development framework inspired by Agile methodology. AgileCoder assigns specialized roles (e.g., Product Manager, Scrum Master, Developer, and Tester) to agents and organizes development into iterative sprints based on user feedback. It further introduces a Dynamic Code Graph Generator to maintain code dependencies and provide relevant context for code generation and modification. Evaluations on HumanEval, MBPP, and ProjectDev show that AgileCoder outperforms existing frameworks such as ChatDev and MetaGPT, demonstrating the effectiveness of integrating Agile principles into multi-agent software development. To provide a broader perspective on designing LLM-based MAS for software engineering tasks, \cite{cai2025designing} examined the quality attributes, design patterns, and design rationale adopted in existing LLM-based MAS. Their analysis shows that \textit{Code Generation} is the most frequently addressed software engineering task, \textit{Functional Suitability} is the most emphasized quality attribute, and \textit{Role-Based Cooperation} and \textit{Self-Reflection} are the most commonly adopted design patterns. They also report that improving code quality and simulating human problem-solving are the primary motivations behind these design choices. Together, these studies demonstrate the effectiveness and growing adoption of LLM-based MAS for software engineering tasks. However, improving their real-world deployment requires a deeper understanding of the challenges practitioners encounter, their underlying causes, and potential solutions.

\subsection{Exploration of Issues, Causes, and Solutions in LLM-based MAS}\label{subsec:MSRinMAS}
While existing studies have demonstrated the potential of LLM-based MAS for software engineering tasks, recent research has started investigating the practical challenges, limitations, and failure patterns of these systems. \cite{liu2025largescale} conducted an empirical study of open-source LLM-based MAS frameworks by analyzing over 42,000 commits and 4,700 resolved issues across eight frameworks. They identified common development and maintenance issues, including bugs, infrastructure problems, and agent coordination challenges, highlighting the difficulties in building reliable MAS. Similarly, \citep{asgari2026} investigated developer challenges in LLM-based MAS frameworks through Stack Overflow discussions and GitHub issues. Their study identified several challenges related to retrieval, agent memory, orchestration, execution control, and runtime reliability, which affect the development and maintenance of LLM-based MAS frameworks. Furthermore, \cite{hamid2026hands} explored practitioner difficulties in agentic frameworks by analyzing 2,658 Stack Overflow posts. They identified challenges related to framework configuration, tool integration, vector database operations, and agent execution behavior, with runtime-related issues emerging as particularly challenging to address. In addition, \cite{cemri2025multi} analyzed over 1,600 execution traces from seven LLM-based MAS frameworks and proposed the MAS Failure Taxonomy (MAST), identifying 14 failure modes across different stages of agent execution. These studies provide insights into LLM-based MAS challenges and failure patterns; however, a comprehensive understanding of practitioner-reported issues, their causes, and solutions remains limited. This study addresses this gap through empirical analysis of the issues of LLM-based MAS.

\subsection{Conclusive Summary}\label{subsec:ConclusiveSummary}
Software repository mining has been widely used to investigate practitioner-reported issues, their root causes, and solutions in software systems, providing a foundation for understanding real-world development challenges \citep{waseem2026microservices}. With the increasing adoption of LLMs in software engineering, researchers have extended repository mining approaches to study issues in LLM-based software systems \citep{cai2025demystifying}. Recently, LLM-based MAS have attracted significant attention by enabling multiple LLM-powered specialized agents to collaboratively perform complex software engineering tasks beyond the capabilities of single-agent LLMs, leading to the development of various frameworks and techniques, such as ChatDev and AgileCoder \citep{qian2024chatdev,nguyen2025agilecoder}. However, relatively few studies have investigated the practical issues, failure patterns, and maintenance challenges of LLM-based MAS \citep{liu2025largescale,cemri2025multi}. Consequently, existing studies on LLM-based MAS provide only limited understanding of the practical problems faced by practitioners, their underlying causes, and the solutions adopted in real-world LLM-based MAS projects. To address this gap, we conducted an empirical analysis of closed GitHub issues from LLM-based MAS projects to derive taxonomies of issues, their underlying causes, and solutions. The comparison with existing studies is summarized in Table~\ref{tab:mas-related-comparison}.

\begin{table}[htbp]
\fontsize{6}{7}\selectfont 
\setlength{\tabcolsep}{2pt} 
\renewcommand{\arraystretch}{1.3} 
\caption{Comparison of empirical studies on LLM-based MAS}
\label{tab:mas-related-comparison}
\raggedright
\begin{tabular}{
>{\raggedright\arraybackslash}m{1.3cm} 
>{\raggedright\arraybackslash}m{1.3cm} 
>{\raggedright\arraybackslash}m{1.4cm} 
m{0.5cm}<{\centering}                  
m{0.8cm}<{\centering}                  
m{0.8cm}<{\centering}                  
m{1cm}<{\centering}}                 
\hline
\textbf{Study} & \textbf{Scope} & \textbf{Data Source} &
\textbf{\makecell{Issues}} & \textbf{\makecell{Causes}} & \textbf{\makecell{Solutions}} & \textbf{\makecell{Mapping}} \\ 
\hline

\cite{He2025MultiAgent} &
LLM-based MAS &
Literature review &
\cmark & \xmark & \xmark & \xmark \\ \hline

\cite{liu2025largescale} &
LLM-based MAS &
GitHub (commits/issues) &
\cmark & \xmark & \xmark & \xmark \\ \hline

\cite{cemri2025multi} &
MAS Failure Patterns &
System logs &
\cmark & \cmark & \xmark & \xmark \\ \hline

\cite{asgari2026} &
AI Agent Frameworks &
Stack Overflow + GitHub &
\cmark & \xmark & \xmark & \xmark \\ \hline

\cite{hamid2026hands} &
AI Agent Frameworks &
Stack Overflow &
\cmark & \xmark & \xmark & \xmark \\ \hline

\textbf{Our Study} &
\textbf{Open-Source LLM-based MAS} &
\textbf{GitHub (944 closed issues)} &
\textbf{\cmark} & \textbf{\cmark} & \textbf{\cmark} & \textbf{\cmark} \\

\hline
\end{tabular}
\end{table}

\section{Conclusions}\label{sec:Conclusions}
In this study, we investigated the issues that practitioners encounter when developing and using open-source LLM-based MAS, as well as their underlying causes and potential solutions. We collected 22,848 closed issues from open-source LLM-based MAS on GitHub that met our inclusion criteria, as shown in Table \ref{tbl:multi-agent-projects}. Then, we applied a mixed automated and manual filtering approach to reduce the dataset to 944 issues related to LLM-based MAS. Finally, we extracted 944 issues, 540 causes, and 427 solutions based on our data extraction criteria. Our results show that \textit{Orchestration \& Execution Issue} is the most common issue faced by practitioners. \textit{Workflow Problem (WP)}, \textit{Tool Integration Problem (TIP)}, and \textit{Memory Problem (MP)} are identified as the most frequent causes of the issues, while \textit{Optimize Workflow (OW)} was employed as the predominant solution to address the issues.

Based on the study results, we provide implications for users, developers, and researchers of open-source LLM-based MAS. Users can address the most frequent \textit{Orchestration \& Execution Issue} through explicit handoff design between agents and multi-agent execution controls, such as retry limits, timeouts, and monitoring mechanisms in LLM-based MAS. They can also refine prompt design to obtain more reliable responses from LLM-based MAS. Developers can improve tool integration reliability through schema validation, execution controls, and tool connection management in LLM-based MAS. Researchers can explore methods to reproduce and diagnose unsolved issues and develop reliable memory management mechanisms to improve context propagation between agents, state synchronization across agents, and memory retrieval for agent interactions in LLM-based MAS.

\section*{Data availability}
The dataset used in this study is available in the replication package \citep{replicationPackage}. 

\section*{Acknowledgments}
This work has been partially supported by the National Natural Science Foundation of China (NSFC) with Grant No. 92582203 and No. 62402348.

\printcredits

\bibliographystyle{cas-model2-names}
\bibliography{references}
\balance
\end{sloppypar}
\end{document}